\documentclass{article}

\usepackage{graphicx}
\usepackage[english]{babel}
\usepackage{amsmath}
\usepackage{amsfonts}
\usepackage{amssymb}
\usepackage{hyperref}
\usepackage{subfig}
\usepackage{float}
\usepackage{mathtools}
\usepackage{csquotes}
\usepackage{overpic}
\usepackage{xcolor}

\usepackage{amsthm}
\theoremstyle{definition}
\newtheorem{lemma}{Lemma}

\newcommand{\edge}[1]{\overset{#1}{\to}}

\let\emptyset\varnothing
\newcommand{\defeq}{\vcentcolon=}
\newcommand{\eqdef}{=\vcentcolon}

\newcommand{\pin}{P_\mathrm{in}}
\newcommand{\pout}{P_\mathrm{out}}
\newcommand{\gin}{G_\mathrm{in}}
\newcommand{\gout}{G_\mathrm{out}}
\newcommand{\zin}{z_\mathrm{in}}
\newcommand{\ptrig}{P_\mathrm{trig}}
\newcommand{\layers}{[L]}
\newcommand{\nicelayers}{[\hat L]}
\newcommand{\pmat}[1]{\begin{pmatrix} #1 \end{pmatrix}}
\renewcommand{\deg}{\mathrm{deg}}
\newcommand{\din}{\deg^{\mathrm{in}}}
\newcommand{\dout}{\deg^{\mathrm{out}}}

\newcommand{\Pois}{\text{Pois}}
\newcommand{\pois}[2]{\text{pois}( #1 ; #2)}
\newcommand{\bin}{\text{bin}}
\newcommand{\inpar}[1]{c^{\text{in}}_{#1}}
\newcommand{\oupar}[1]{c^{\text{out}}_{#1}}

\usepackage[sorting=none]{biblatex}
\usepackage{xr} 

\let\ticklabelsize\tiny
\definecolor{ForestGreen}{RGB}{0, 124, 48}

\title{Cascades on Networks with Functional Structure}
\author{Christian Kluge\footnote{Technical University of Munich, School of Computation Information and Technology, Department of Mathematics, Garching b.~M\"unchen, Germany} and Christian Kuehn\footnotemark[\value{footnote}]~\footnote{Munich Data Science Institute, Garching b.~M\"unchen, Germany}~\footnote{Complexity Science Hub Vienna, Vienna, Austria}}

\begin{document}
\maketitle
\begin{abstract}
    We consider a version of the Watts threshold model on directed multiplex configuration model networks,
    and present a detailed analysis of the cascade size, single-seed cascade probability and cascade condition.
    We then introduce a smaller class of network models that we call \textit{constrained multiplex networks}, which is designed to represent networks with so-called \textit{functional} or \textit{complementary} structure.
    We find that the particular choice of functional structure affects the phase transitions of the cascade model in a variety of ways.
\end{abstract}


\section{Introduction}
Complex contagion on networks has been an active area of research in the past twenty years, with applications ranging from the spreading of behavior in social networks \cite{Centola2010} to systemic risk in financial networks \cite{Battiston2018}.
The Watts Threshold Model (WTM) \cite{Watts2002} - also called the Linear Threshold Model or Watts' Cascade Model - has been a classic in this area.
It is one of a variety of spreading processes used to examine the fragility or robustness of networks.
For a recent overview of this broader context, we refer the reader to \cite{Artime2024}.
In the WTM, nodes are activated once a sufficiently large fraction of their neighbors are active. The precise value of \enquote{sufficiently large} is given by a \textit{threshold fraction} $\phi\in[0,1]$, which may be random and different for each node.
The WTM is often presented as modeling the spreading of beliefs or behaviors in a social network.
In this interpretation, nodes correspond to individuals and the \enquote{active} state means the individual has adopted the belief or behavior.
Initially, the model was analyzed on treelike configuration-model networks \cite{Gleeson2007}, but its analysis and that of similar models has also been carried out for clustered random networks \cite{Hackett2011, Hackett2013, Zhuang2017, Keating2022}, which is natural as social networks feature high clustering \cite{Newman2003a}.
Separate from the WTMs interpretation as a social process, it has also found use as a simple, analytically tractable model of default cascades among banks \cite{Gai2010,Burkholz2016,Amini2016}, with the goal of better understanding financial systemic risk.
In this context, nodes are banks and \enquote{active} means the bank has defaulted.

In recent years, there have been efforts to extend the modeling of systemic risk beyond the financial sector to the real economy \cite{Inoue2019,Diem2022}, sparked by the availability of detailed, firm-level supply chain data \cite{Pichler2023}.
We believe that the WTM can - in appropriately modified form - play a role in this process.
This is because stylized, analytically tractable models allow us to develop an intuitive understanding for the mechanisms at play, even if they lack the quantitative conclusions that more complex agent-based models can provide.
This would require analyzing a variant of the WTM on networks capable of representing supply chains, which are inherently directed and multiplex, as one needs to distinguish suppliers and customers, as well as different goods.

The structure of supply chain networks is quite interesting in its own right, as it has recently been proposed that they feature so-called \enquote{complementarity-driven} or \enquote{functional} structure \cite{Mattsson2021}.
This relatively new concept originates from work on protein interaction networks \cite{Kovacs2019}.
It is in some sense the opposite of the principle of homophily - the tendency of nodes to connect to others that are similar to themselves -, which is well-known from social networks \cite{McPherson2001}.
By contrast, in networks driven by complementarity, nodes connect to others that are different and in some way compatible.
In the case of supply chains, this means that we expect a link to connect a producer of some good to a consumer of that good, but not two producers or consumers.
As complementarity is a new concept that is still being developed, not much is currently known about its influence on network dynamics.
An appropriate network model to explore this interplay of complementarity and dynamics appears to be the multiplex configuration model, which has been found capable of reproducing the complementary structure of supply chains \cite{Mattsson2021}.

So far, three extensions of the WTM to multiplex networks have been proposed.
The first is the \textit{or}-rule \cite{Brummitt2012}, in which the neighborhoods on each layer are considered separately.
A node activates if it has a sufficient fraction of active neighbors on \textit{any one} layer.
Second, in the \textit{and}-rule \cite{Lee2014} the neighborhoods on each layer are also considered separately, but a node needs enough active neighbors on \textit{every} layer to activate.
In the third rule \cite{Yagan2012}, the numbers of active neighbors on each layer are combined by an average that may give different weights to the contribution of different layers. Nodes activate when this aggregate value exceeds the threshold.
Of these, only the \textit{or}-rule appears suitable for modeling firm-level supply chain contagion.

Specifically, we would consider an \textit{active} node to represent a \textit{failed} firm, i.e.~a firm that has been impacted by the supply shock and is unable to adequately supply its customers.
The \textit{or}-rule then means that different input goods (layers) cannot be substituted for each other, and the complete loss of a single input is guaranteed to cause a firm to fail.

Though there is much existing work on multiplex WTM variants, few of the results can be immediately adapted to our purposes. To our knowledge, there is no comprehensive analysis in the literature of the \textit{or}-rule on directed networks.
And while there has been work on the effects that more explicit network structure can have on cascade dynamics \cite{Zhuang2017,Unicomb2019}, it was not aimed at complementarity.

In this paper, we analyze the or-rule threshold model on general multiplex configuration model networks with the aim of understanding the impact of functional structure on the cascade dynamics.
As the multiplex configuration model in general has an immense number of degrees of freedom, we also introduce a smaller class of networks that we call \textit{constrained multiplex networks}.
This subclass is tailored to make the functional structure of the network explicit, and it also incorporates some constraints on the network structure that are natural for supply chains.
We give examples of a variety of phenomena that can be induced by this network structure, such as the appearance of additional phase transitions not typically seen in the Watts threshold model, as well as a change in the order of a typical one.
Additionally, we observe that the usual method for computing the cascade size may fail for single-seed cascades in certain networks.
Despite our motivation originating partly from the study of supply chains and systemic risk, we will keep our approach general, as the topic of functional structure is of wider interest.

The remainder of this paper is structured as follows:
In section \ref{sec-def}, we define the network models and the dynamics.
We then proceed with the analysis of cascades starting from infinite seeds in section \ref{sec-size} and finite seeds in section \ref{sec-prob}.
Finally in section \ref{sec-constraint-examples}, we present some examples of different network structures and their impact on cascades.


\section{Notation \& Model Definitions}
\label{sec-def}
A \textit{multiplex network} is a multigraph in which each link is labeled with a \textit{type} from some set, which we will take to be $[L]\defeq\{1,\dots,L\}$ for some integer $L\ge1$.
We will denote the types by $\alpha, \beta, \gamma \in \layers$.
All networks in this work will be directed, and to denote a type-$\alpha$-link from node $u$ to node $v$, we will write $u\edge{\alpha}v$.
The collections of all links of the same type are referred to as the \textit{layers} of the network.
We will use the terms \enquote{layer} and \enquote{type} interchangeably, thus we also call $\layers$ the \textit{layer set}.
We let $\mathbb{N}$ denote the natural numbers including $0$.
Regarding node degrees, we write the in-degree of a node $v$ as $\din(v)=J\in\mathbb{N}^L$ and the out-degree as $\dout(v)=K\in\mathbb{N}^L$,
where the components of $J$ and $K$ count links of each type separately.
So given a node with full degree $\deg(v)=(J,K)\in\mathbb{N}^L\times\mathbb{N}^L$, we denote its in-degree on some layer $\alpha\in\layers$ by $\din_\alpha(v)=J_\alpha\in\mathbb{N}$.

\subsection{Cascade Model}
Each node $v$ has a state $\eta_t(v)\in\{0,1\}$ at every time step $t\in\mathbb{N}$. We call nodes with $\eta_t(v)=1$ \textit{active} and those with $\eta_t(v)=0$ \textit{inactive}.
Additionally, we also call a link $u\edge{\alpha}v$ \textit{active} if its starting node $u$ is active.
This shorthand will prove convenient throughout our work.
The update rule is simple:
Active nodes remain active forever, and inactive nodes become active if - on at least one layer where they have in-neighbors - all their in-neighbors are active.
Thus, if a node has no in-neighbors on some layer, the node does \textit{not} automatically activate.
This way of handling the edge case is motivated by seeing the links as pathways for the spreading of cascades:
The \textit{complete absence} of such pathways should not propagate the cascade.
In summary, we have $\eta_{t+1}(v)=1$ if $\eta_t(v)=1$ or if there exists $\alpha\in \layers$ such that
\begin{equation}
    \{u\mid u\edge{\alpha}v\}\ne\emptyset \ \text{and}\ \forall u\in\{u\mid u\edge{\alpha}v\}: \eta_t(u)=1.
\end{equation}
Compared to typical studies of threshold models, this means that we fix all thresholds to $\phi=1$.
We do this for two reasons: First, we wish to limit the number of control parameters under consideration.
The network models introduced in the next sections already have more parameters than we can explore comprehensively.
Second, the multiplex degree distribution already gives us a way to tune the individual robustness of nodes: A node is hard to activate if all its in-links are on the same layer - but if they are spread across many different layers, the node activates very easily.

To initialize the model, some nodes are already in the active state at time $0$.
These are referred to as \textit{seed nodes}.
In this work, we will examine both macroscopic seeds and microscopic seeds.
In the case of a macroscopic seed, the initial node states are i.i.d.~random variables.
Each node is active with probability $\rho_0$, and inactive with probability $1-\rho_0$,
where $\rho_0\in[0,1]$ is a parameter referred to as the \textit{seed size}.
The absolute number of seed nodes is therefore dependent on the total number of nodes in the network.
For microscopic seeds, a fixed number of seed nodes is chosen uniformly at random, and this number is independent of the size of the network.

\begin{figure}[t]
    \centering
    \subfloat{\begin{overpic}[width=0.23\linewidth]{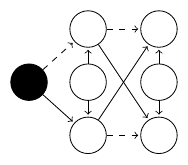}
        \put(3,88){(a)}\end{overpic}}
    \subfloat{\begin{overpic}[width=0.23\linewidth]{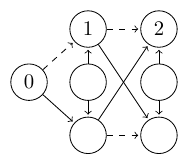}
        \put(3,88){(b)}\end{overpic}}\\
    \subfloat{\begin{overpic}[width=0.23\linewidth]{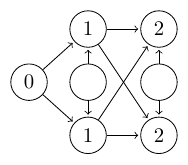}
        \put(3,88){(c) All $\phi\in[0,\frac{1}{2}]$.}\end{overpic}}
    \subfloat{\begin{overpic}[width=0.23\linewidth]{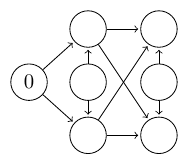}
        \put(3,88){(d) All $\phi\in(\frac{1}{2},1]$.}\end{overpic}}
    \subfloat{\begin{overpic}[width=0.23\linewidth]{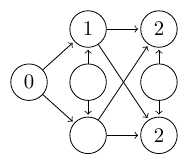}
        \put(3,88){(e)}
        \put(32,85){\scriptsize $\phi = \frac{1}{2}$}
        \put(32,-2){\scriptsize $\phi = 1$}
        \put(72,85){\scriptsize $\phi = \frac{1}{3}$}
        \put(72,-2){\scriptsize $\phi = \frac{1}{3}$}
    \end{overpic}}
    \caption{Example of the multiplex update rule, in comparison to the classic Watts model.
    The network is shown in (a), with the seed node filled in. The dashed arrows indicate layer-$2$-links, the solid links are on layer $1$.
    (b) shows how the cascade plays out under the update rule considered in this paper. The number in each node shows the time step at which that node activates. Nodes that never activate are left empty.
    (c-e) show the cascade if we ignore the different link types and apply the classic Watts model instead,
    with different thresholds $\phi$.
    In (e), the threshold of each node is chosen to match the smallest number of neighbors that could activate it in the multiplex update rule.
    }
    \label{fig:example-modeldef}
\end{figure}

An application of the update rule to a small example network is illustrated in Fig.~\ref{fig:example-modeldef}, where we also see that its behavior cannot be exactly reproduced by the standard Watts model on a monoplex version of the network.

\subsection{General Network Model}
As is commonly done, we will use an asymptotic network model to investigate the cascade behavior in the limit of infinitely large sparse networks, and we will not consider finite-size effects in our analysis.
Our network model is a straightforward extension of the well-known \textit{random graphs with arbitrary degree distribution} \cite{Newman2001} (or \textit{configuration model}) to directed multiplex networks.

A random network in our model is specified by the set $\layers$ of layer indices, and a probability mass function $P:\mathbb{N}^L\times\mathbb{N}^L \to [0,1]$ giving the joint distribution of multiplex in- and out-degrees.
The probability of a node having in-degrees $J$ and out-degrees $K$ is written as $P(J,K)$.

Since $P$ is a $2 L$-dimensional multivariate distribution, it has a large number of degrees of freedom.
While this gives the network model considerable expressive power, it also makes the general results we will derive in terms of $P$ difficult to grasp intuitively.
To alleviate this, we will repeatedly reference the following simple example:
By a \textit{network with i.i.d.~degrees}, we refer to the special case of a degree distribution $P$ satisfying
\begin{equation}
    P(J,K) = \prod_{\alpha\in \layers} \pin(J_\alpha) \pout(K_\alpha),
\end{equation}
where $\pin:\mathbb{N}\to [0,1]$ and $\pout:\mathbb{N}\to [0,1]$ are probability mass functions of univariate distributions of in- and out-degrees, respectively.

\subsection{Constrained Multiplex Network Model}
\begin{figure}
    \centering
        \begin{overpic}[width=.65\textwidth]{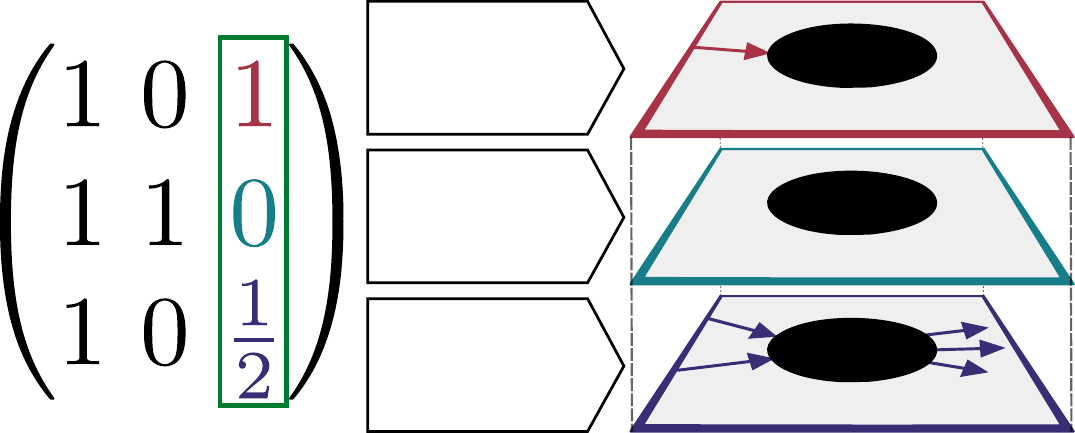}%
            \put(35,34.5){\small Guaranteed}%
            \put(35,30.5){\small in-Links}%
            \put(35,19){\small No in-Links}%
            \put(35,7){\small $50\%$ Chance}%
            \put(35,3){\small of in-Links}%
            \put(0,42){\small Constraint Matrix}%
            \put(59,42){\small Example {\color{ForestGreen}Type-3} Node}%
        \end{overpic}
    \caption{Illustration showing how node degrees are constructed in a three-layer constrained multiplex network.
    Note that since the example depicts a type-$3$-node, its out-stubs are all on layer $3$, and the configuration of its in-stubs is given by the third column of the constraint matrix.}
    \label{fig:illustration-constraints}
\end{figure}

The \textit{constrained multiplex network} is our attempt at modeling complementarity-driven networks while keeping the number of parameters manageable.
In addition, we want to cleanly separate the marginal degree distributions and the link density from the functional structure.
Our idea is to assign a \enquote{role} to each node, and to only allow links between nodes with compatible roles.
Interpreted in terms of modeling supply chains, this role would correspond to the business model of a firm, represented e.g.~by its industry classification.
In practice, we identify these \enquote{roles} with the link types $[L]$, and we only allow nodes to have out-links of the type matching their role.
In supply chain modeling, this would correspond to assuming a perfect correspondence of industry classes to product classes.
The in-link types are constrained by specifying a \textit{constraint matrix} $C\in[0,1]^{L\times L}$, where the entry $C_{\beta\alpha}$ is the probability that a type-$\alpha$-node is allowed to have in-links of type $\beta$.
So if e.g.~$C_{\beta\alpha}=0$, then no type-$\alpha$-node may have an incoming type-$\beta$-link,
if $C_{\beta\alpha}=\frac{1}{2}$, then half of the type-$\alpha$-nodes may have incoming type-$\beta$-links, and so on.
In the supply chain interpretation, this means that a good is always produced from the same inputs, with some random variation.
The actual number of incoming type-$\beta$-links at each node is then sampled independently from a univariate in-degree distribution $\pin$, which we are free to choose.
Note that if $\pin(0)$ is positive, the actual number of type-$\alpha$-nodes with incoming type-$\beta$-links will be lower than $C_{\beta\alpha}$.
To keep things simple, we use the same in-degree distribution for all link types.
Analogously, we also choose a single out-degree distribution $\pout$ for all nodes.
See Fig.~\ref{fig:illustration-constraints} for an illustration.

The precise construction is as follows: The model parameters are a constraint matrix $C\in[0,1]^{L\times L}$, a univariate in-degree distribution $\pin$, and a family of univariate out-degree distributions $(\pout^z)_{z>0}$ s.t.~$\pout^z$ has mean $z$.
Each node is assigned a type from $\layers$ uniformly at random.
Let $\delta_0:\mathbb{N}\to\{0,1\}$ denote the probability mass function of the Dirac measure at $0$.
Then the in-degree-component $J_\beta$ of a type-$\alpha$-node is distributed according to the mixture distribution
\begin{equation}
    (1-C_{\beta\alpha})\cdot\delta_0+C_{\beta\alpha}\cdot\pin,
\end{equation}
i.e.~$J_\beta$ is zero with probability $1-C_{\beta\alpha}$ and otherwise drawn from $\pin$. The components of the in-degree vector $J$ are conditionally independent given the type of the node.
Thus the mean in-degrees $(z_\alpha)_{\alpha\in\layers}$ of all layers are already determined by $C$ and $\pin$.
Specifically, let $\zin$ be the mean of $\pin$ and obtain
\begin{equation}
    z_\alpha= \frac{\zin}{L}\sum_{\beta\in\layers} C_{\alpha\beta}.
\end{equation}
In order to match the in- and out-degrees on each layer, let the out-degree component $K_\alpha$ of a type-$\alpha$-node be distributed according to $\pout^{Lz_\alpha}$, independently of all in-degrees.
All other out-degree components $K_\beta, \beta\ne\alpha$ are almost surely $0$.

Putting all of this together, the overall degree distribution in a constrained multiplex network is given by
\begin{equation}\label{eq-p-constrained}
\begin{split}
P(J,K) = \frac{1}{L} \sum_{\alpha\in\layers}
    \Bigg[
        \pout^{Lz_\alpha}(K_\alpha)
        &\left( \prod_{\beta\ne\alpha}\delta_0(K_\beta) \right)\\
        \cdot &\left( \prod_{\beta\in\layers} (1-C_{\beta\alpha})\delta_0(J_\beta) + C_{\beta\alpha}\pin(J_\beta) \right)
        \Bigg].
\end{split}
\end{equation}


\section{Analysis: Macroscopic Seeds}
\label{sec-size}
We begin by analyzing the case of macroscopic seeds.
In this section, we first find equations whose iteration yields the development of the expected cascade size through time.
From these equations, we then derive a condition for the emergence of macroscopic cascades from microscopic seeds, and finally use a simple example to provide some intuition about these findings.

\subsection{Expected Cascade Size}
We initialize our model with some seed size $\rho_0>0$, i.e.~each node is independently active at time $0$ with probability $\rho_0$, and inactive with probability $1-\rho_0$.
We are interested in the expected cascade size $\rho(t)$, i.e.~the expected fraction of nodes that is in the active state at each time $t\ge 0$.

We use an approach sometimes referred to as \enquote{message-passing}, which is a common technique to find cascade sizes in threshold models (see for example \cite{Gleeson2007,Brummitt2012,Yagan2012,Shrestha2014}) and is often traced back to \cite{Dhar1997}, who
used the idea to derive a self-consistency equation for the average magnetization in the random-field Ising model on the Bethe lattice.
A good introduction to the method and its applications can be found in \cite{Newman2023}.

To begin, we note that
$\rho(t)$ is equal to the probability that a randomly chosen node is active at time $t$.
Fix a random node $v$ and some time $t$. With probability $\rho_0$, $v$ is a seed node and is therefore active at all times, including $t+1$.
With probability $1-\rho_0$, $v$ is not a seed node and is only active at time $t+1$ if it was activated via its in-links.
Since active nodes never deactivate, this is equivalent to the existence of
a layer $\alpha$ where $v$ has incoming links and all these links are active at time $t$.
Since the network model we are considering is locally tree-like \cite{Soederberg2003a}, we treat all incoming layer-$\alpha$-links as statistically identical and independent.
It thus suffices to know the probability $q_\alpha(t)$ that a randomly chosen layer-$\alpha$-link is active at time $t$, and we can write
\begin{equation}
    \rho(t+1) = \rho_0+(1-\rho_0)\sum_{J,K\in\mathbb{N}^L} P(J,K)\left( 1- \prod_{\overset{\alpha\in \layers}{J_\alpha>0}}\left( 1- q_\alpha(t)^{J_\alpha} \right)\right).
\end{equation}
Next, we determine $q_\alpha(t)$.
Fix a random layer-$\alpha$-link $u\edge{\alpha}v$.
By definition, this link is active if its source node $u$ is active.
The reasoning for $u$ is the same as for the focal node $v$, but the degree distribution of $u$ must account for the fact that links are more likely to lead to high-degree nodes:
The probability that $u$ has degree $(J,K)$ is equal to $\frac{1}{z_\alpha}P(J,K) K_\alpha$, where $z_\alpha = \sum_{J,K}P(J,K)K_\alpha$ is the mean (in- or out-) degree on layer $\alpha$.
We obtain
\begin{equation}
    q_\alpha(t+1) = \rho_0+(1-\rho_0)\sum_{J,K}\frac{P(J,K)K_\alpha}{z_\alpha}\left( 1- \prod_{\overset{\beta\in \layers}{J_\beta>0}}\left( 1- q_\beta(t)^{J_\beta} \right)\right).
\end{equation}
Using straightforward algebra, we rewrite both of these equations as
\begin{equation}\label{eq-cascadesize-node}
    \rho(t+1)=1- (1-\rho_0) \sum_{J,K} P(J,K)\prod_{\overset{\alpha\in \layers}{J_\alpha>0}} \left( 1-q_\alpha(t)^{J_\alpha} \right)
\end{equation}
and
\begin{equation}\label{eq-cascadesize-edge}
     q_\alpha(t+1)=1- \frac{1-\rho_0}{z_\alpha} \sum_{J,K} P(J,K) K_\alpha \prod_{\overset{\beta\in \layers}{J_\beta>0}}\left( 1- q_\beta(t)^{J_\beta} \right).
\end{equation}
To determine the development of the cascade up to some final time $T\in\mathbb{N}$,
we initialize $\rho(0)=q_\alpha(0)=\rho_0$ for all $\alpha\in\layers$, and iterate \eqref{eq-cascadesize-edge} to compute $q_\alpha(t)$ for all $\alpha\in\layers$ and all time steps $0<t<T$.
We then use \eqref{eq-cascadesize-node} to obtain the expected cascade size $\rho(t)$ for all time steps $0<t\le T$.
In general, these computations must be performed numerically.
However, one can still determine some properties of the iteration \eqref{eq-cascadesize-edge} analytically, as we will see in the next section.

\subsection{Cascade Condition}
\label{sec-size-cond}
A natural next step is to derive a \textit{cascade condition} \cite{Watts2002,Gleeson2007},
identifying those parameter regions in which macroscopic cascades may be set in motion by the activation of a single node.
This corresponds to a seed size that is vanishingly small compared to the total number of nodes, i.e.~$\rho_0 = 0$.
In this case, the point where all $q_\alpha=0$ is a fixed point of equation \eqref{eq-cascadesize-edge}.
We expect macroscopic cascades to occur if and only if this fixed point is locally asymptotically unstable.

We let $\rho_0=0$ and write equation \eqref{eq-cascadesize-edge} as $q(t+1) = f(q(t))$, where $f:[0,1]^{L}\to [0,1]^{L}$ is given by
\begin{equation}
    f_\alpha(q) \defeq 1- \frac{1}{z_\alpha} \sum_{J,K} P(J,K) K_\alpha \prod_{\overset{\beta\in \layers}{J_\beta>0}}\left( 1- q_\beta^{J_\beta} \right).
\end{equation}
We want to determine stability of the fixed point $q^\star \defeq (0,\dots,0)\in [0,1]^L$ and find
\begin{equation}
    \frac{\textnormal{d}f_\alpha}{\textnormal{d}q_\gamma}(q) = \frac{1}{z_\alpha} \sum_{\overset{J,K}{J_\gamma>0}} P(J,K) K_\alpha \cdot J_\gamma q_\gamma^{J_\gamma-1} \cdot \prod_{\overset{\beta\in \layers \setminus \{\gamma\}}{J_\beta>0}}\left( 1- q_\beta^{J_\beta} \right),
\end{equation}
which for $q\to q^\star$ converges to
\begin{equation}\label{eq-condition-matrix-macro}
    \frac{1}{z_\alpha} \sum_{\overset{J,K}{J_\gamma=1}} P(J,K) K_\alpha \ {=\vcentcolon}\ A_{\alpha\gamma}.
\end{equation}
Thus, in order to determine the possibility of single-seed cascades, one merely needs to compute the dominant eigenvalue of the matrix $A\in \mathbb{R}^{L\times L}$ with entries $(A_{\alpha\beta})_{\alpha,\beta\in \layers}$.
If its absolute value exceeds $1$, single-seed cascades are possible.

\subsection{Simple Special Case}
\label{sec-size-iid}
While the results derived above are quite general, they are also difficult to interpret.
In this section, we consider the special case of i.i.d.~degrees to illustrate some behaviors of the system.
We assume
\begin{equation}
    P(J,K)=\prod_{\alpha\in \layers} \pin(J_\alpha)\pout(K_\alpha).
\end{equation}
In this case, all layers have identical structure and correspondingly, equation \eqref{eq-cascadesize-edge} no longer depends on $\alpha$.
Furthermore, it simplifies to be the same as \eqref{eq-cascadesize-node}:
Since we assume in- and out-degrees to be independent, the in-neighbors of our focal node have the same in-degree distribution as the focal node itself.
As only in-degrees matter for our activation rule, the in-neighbors are thus active with the same probability as the focal node, eliminating the need for a separate equation.
We arrive at the single recursive equation (see Appendix \ref{apx-size-iid} for the derivation)
\begin{equation}\label{eq-cascadesize-iid}
    \rho(t+1) = 1-(1-\rho_0)\cdot \Big( 1 + \gin(0) -\gin(\rho(t))  \Big) ^{L},
\end{equation}
where $\gin$ is the probability generating function of $\pin$.

In equation \eqref{eq-cascadesize-iid} we can see some interesting properties of the system:
Since $\gin$ is a probability generating function, we always have $\gin(0)\le\gin(x)\le 1$ for $x\in[0,1]$.
The second term thus shrinks as $L$ increases, pushing the value of $q$ closer to $1$.
The interpretation is clear: Increasing the number of layers - while keeping the degree distribution on each layer fixed - will increase the size of cascades, in line with previous findings \cite{Brummitt2012}.
Additionally, we can see that a special role is taken by $\gin(0) = \pin(0)$, which is the probability of a node having no incoming links on a particular layer.
Such a node is immune to activation on that layer and thus large values of $\pin(0)$ will inhibit the development of cascades.
On the topic of the degree distribution, we also notice that $\pout$ is entirely absent from equation \eqref{eq-cascadesize-iid}.
This is because we assumed the in- and out-degrees of each node to be independent.
If there was a correlation between in- and out-degrees (as there often is in real networks), then $\pout$ would affect the expected cascade size.
The same principle applies to single-layer networks, where it has been noted that even when the in- and out-degrees are independent, $\pout$ may still affect the distribution (as opposed to merely the expectation) of cascade sizes on finite networks \cite{Gleeson2008}.

We now turn to the cascade condition, which hinges on the leading eigenvalue of the matrix $A$ with entries
\begin{equation}
    A_{\alpha\beta} = \frac{1}{z_\alpha} \sum_{\overset{J,K}{J_\beta=1}} P(J,K) K_\alpha,
\end{equation}
which simplifies to $A_{\alpha\beta}=\pin(1)$ in our special case (see Appendix \ref{apx-cond-iid}).
Thus, the relevant eigenvalue of $A$ is $\lambda=L\cdot\pin(1)$ and we obtain the simple cascade condition
\begin{equation}
    L\cdot\pin(1) > 1.
\end{equation}

We have already seen that the number of layers increases the size of cascades, so it is natural to see it again here.
The other component is $\pin(1)$, which is the probability that a random node has exactly $1$ in-link on a given layer.
Such nodes may be activated by a single neighbor and are therefore especially vulnerable to cascades.
It is well known that these single-neighbor-activations are the pathway by which microscopic seeds lead to macroscopic cascades \cite{Watts2002}, so the importance of $\pin(1)$ is expected.

It may be surprising that these are the only two factors that appear.
As previously stated, the absence of $\pout$ is due to the assumed independence of in- and out- degrees.
But the absence of all other parts of $\pin$ is notable, as it indicates
that the possibility of cascades is in particular unaffected by the tail of the degree distribution.
This is because high-in-degree nodes are immune to single-neighbor-activations, and thus their precise degree distribution only becomes relevant once the cascade has grown to a positive fraction of the network.
Contrast this with overload cascades \cite{Motter2002} and some other systems considered in network science (e.g.~attack tolerance \cite{Albert2000} or epidemic spreading \cite{PastorSatorras2001}), where it has long been known that heavy-tailed degree distributions greatly impact the dynamics.

\subsection{In Constrained Multiplex Networks}
We now transfer our analytical results to the special case of constrained multiplex networks.
We only state the resulting equations, the details of the derivations can be found in Appendix \ref{apx-constrained-derivations}.

Beginning with the expected cascade size,
we substitute equation \eqref{eq-p-constrained} into equations \eqref{eq-cascadesize-node} and \eqref{eq-cascadesize-edge}, and obtain
\begin{align} \label{eq-cascadesize-node-constrained}
\rho(t+1)
    &= \frac{1}{L} \sum_{\alpha\in\layers} q_\alpha(t+1)\\
q_\alpha(t+1)
    &= 1 - (1-\rho_0)\prod_{\beta\in\layers} 
        1+C_{\beta\alpha}\cdot\Big(\gin(0)-\gin(q_\beta(t))\Big). \label{eq-cascadesize-edge-constrained}
\end{align}
For the cascade condition we similarly substitute \eqref{eq-p-constrained} into \eqref{eq-condition-matrix-macro} and find
\begin{equation}
    A_{\alpha\gamma} = C_{\gamma\alpha} \cdot \pin(1).
\end{equation}
Recall that the emergence of large cascades hinges on whether the dominant eigenvalue of $A$ exceeds $1$ in absolute value.
As we can see, the eigenvalues of $A$ are simply those of $C$, scaled by $\pin(1)$.
Letting $\lambda_C$ denote the dominant eigenvalue of $C$, cascades are therefore possible if and only if
\begin{equation}\label{eq-condition-constrained}
    |\lambda_C|\cdot \pin(1) > 1.
\end{equation}
Since $C$ and $\pin$ are explicit parameters to the constrained network model, this makes it easy to choose network structures that are guaranteed to have (or not have) cascades, or that are close to phase transitions.


\section{Analysis: Microscopic Seeds}
\label{sec-prob}
With the above results, we can determine whether large cascades could be triggered by a vanishing number of nodes and - if so - what size we expect them to reach.
But even when cascades from microscopic seeds are possible, they are far from guaranteed.
After all, we could always choose poorly connected seed nodes by pure chance.
In this section, we derive an iterated map that allows us to determine the probability of a cascade.
From this, we then derive an alternative cascade condition that is provably equivalent to the one in section \ref{sec-size-cond}.
We again illustrate our analysis with a simple example.

\subsection{Cascade Probability}
By the \textit{single-seed cascade probability} $\ptrig$, we mean the probability that a single seed node triggers a cascade that reaches a nonzero fraction of the network - in the limit of infinite network size.
This problem is usually treated \cite{Watts2002, Gleeson2008a, Yagan2012} by considering so-called \textit{vulnerable nodes}, which may be activated by a single active neighbor.
If these nodes form an infinite connected component (referred to as \textit{the vulnerable cluster}), then the activation of a single node in the cluster will lead to the activation of the entire, infinite cluster.
It is typically assumed that the cascade then proceeds as it does in the case of a macroscopic seed,
though we will see an example in section \ref{sec-example-nested} where this is not the case.

In the case of directed multiplex networks, this approach becomes significantly more complicated.
While Yağan and Gligor \cite{Yagan2012} have managed to transfer the concept of a vulnerable cluster to multiplex networks, they ultimately derived the cascade probability in their model without relying on clusters, using a branching process argument instead.

Our approach will be similar, and we will not speak of vulnerable nodes or clusters. Instead, we introduce the concept of \textit{vulnerable links}.
We call a link on layer $\alpha$ \textit{vulnerable} if its target node has no other incoming layer-$\alpha$-links.
Thus, the cascade is guaranteed to propagate across such links.
And we know that -- when the cascade is still small relative to the network -- it will \textit{only} propagate along such vulnerable links \cite{Watts2002}.
This is because our random networks are locally treelike \cite{Soederberg2003a} and the probability for any two active nodes to have a common neighbor vanishes.

With this in mind, we now consider the early stages of a cascade exclusively in terms of the vulnerable links.
Recall that a link is called \textit{active} if its starting node is active.
An active vulnerable link will activate all vulnerable links that start at its endpoint.
This is essentially a branching process, where the active vulnerable links are the particles and the number of offspring of a particle is drawn based on the degree distribution of the network.
Further, this branching process is multi-type: We need to distinguish between links on different layers, since the degree distributions of their target nodes are different.

A fundamental property of branching processes is the transience of non-zero states, meaning that the process will either die out or it will diverge to infinity \cite{Athreya1972}.
For us, the former case corresponds to a cascade that terminates at a finite size,
while the latter case corresponds to a cascade that reaches a macroscopic fraction of the network.
Our plan is to compute the extinction probability of the branching process, and thus obtain $\ptrig$.

However, there is a complication:
To prove the transience of nonzero states for multi-type branching processes, one requires the branching process to be \textit{positive regular} and \textit{nonsingular} (see section II.6 of \cite{Harris1963}).
Both assumptions may be violated by our cascade process, depending on the degree distribution of the network.

A branching process is called \textit{singular} if every particle is guaranteed to have exactly one offspring.
In our setting, this corresponds to a degree distribution such that each node has exactly one outgoing link on a layer where all links are vulnerable. All other outgoing links must be on layers where no links are vulnerable.
While it easy to construct examples of networks with such degree distributions, it seems unlikely that a realistic network arising in applications would have this property.
From now on, we will assume our networks to be such that the resulting branching process is nonsingular.
We do not believe this to be a serious restriction.

The definition of positive regularity is slightly more involved.
Let $M$ be the \textit{mean matrix} of the branching process, whose entries $M_{ij}$ are the expected number of type-$j$-offspring of a single type-$i$-particle in one generation.
The process is called \textit{positive regular} if there exists $n\in\mathbb{N}$ such that all entries of $M^n$ are strictly positive.
It is not immediately clear what restrictions this imposes on the degree distribution in general,
but in section \ref{sec-prob-constrained} we will see that this requirement becomes more concrete for constrained multiplex networks.

There is one obvious consequence of positive regularity that we want to address here:
In order for the resulting process to be positively regular, the network must not have any layers without vulnerable links.
Fortunately, we are able to simply exclude such layers from our branching process considerations, since they do not matter in the early stages of a cascade.
To this end, let $\hat L$ be the cardinality of
\begin{equation}
    \left\{ \alpha\in \layers : \sum_{\overset{J,K}{J_\alpha=1}} P(J,K) > 0 \right\} \subset\layers,
\end{equation}
the subset of layers that contain vulnerable links.
Without loss of generality (reordering the labels if need be), this set is exactly $\nicelayers \defeq \{1,\dots,\hat L\}$.
Then let $\hat P: \mathbb{N}^{\hat L} \times \mathbb{N}^{\hat L} \to [0,1]$ be the probability mass function of the marginal degree distribution on those layers.

From now on, we require the network \textit{restricted to the layers} $\nicelayers$ to be such that the resulting process is positively regular.
This is a significant restriction on the generality of networks that we are able to consider, but it is necessary to apply branching process theory.

It may seem unusual to put so much emphasis on the assumptions behind the branching process analysis, particularly as these considerations are usually absent from other works studying threshold models on networks.
This is because the issue does not arise if one studies single-layer networks or multiplex networks with independent layers, or if one focuses only on the macroscopic seed case.
However, we want to study precisely the consequences of dependencies between the different layers of a multiplex network. So we must keep in mind that some of the assumptions underpinning our analysis may be violated even if the network under consideration is not obviously pathological.
In section \ref{sec-example-nested} we will see a simple example of a network that violates positive regularity.

To compute the extinction probability, we first need the generating functions for the offspring distributions of each particle type.
In our case, this translates into the number of vulnerable links on each layer that start at a node reached by following a random vulnerable layer-$\alpha$-link.
Let $\alpha \in \nicelayers$ and let
\begin{equation}\label{eq-cascadeprob-v}
    v_\alpha = \frac{1}{z_\alpha} \sum_{\overset{J,K \in\mathbb{N}^{\hat L}}{J_\alpha=1}} \hat P(J,K)
\end{equation}
be the probability that a random layer-$\alpha$-link is vulnerable.
The probability generating function of the out-degree of the destination of a vulnerable layer-$\alpha$-link is
\begin{equation}
    g_\alpha(\boldsymbol{x}) = \frac{1}{v_\alpha z_\alpha} \sum_{\overset{J,K\in\mathbb{N}^{\hat L}}{J_\alpha=1}} \hat P(J,K) \prod_{\beta\in\nicelayers} x_\beta^{K_\beta}
\end{equation}
and the generating function of the number of outgoing vulnerable links is
\begin{equation}\label{eq-cascadeprob-offspring}
    G_\alpha(\boldsymbol{x}) = \frac{1}{v_\alpha z_\alpha} \sum_{\overset{J,K\in\mathbb{N}^{\hat L}}{J_\alpha=1}} \hat P(J,K) \prod_{\beta\in\nicelayers} \big( (1-v_\beta) +v_\beta x_\beta \big)^{K_\beta}.
\end{equation}
These $G_\alpha$ are the relevant offspring generating functions.
We now combine them into one single function $G:\mathbb{R}^{\hat L}\to \mathbb{R}^{\hat L}$ defined by $(G(\boldsymbol{x}))_\alpha = G_\alpha(\boldsymbol{x})$.
A standard result on branching processes (see e.g.~Thm.~2 in Chapter V, Section 3 of \cite{Athreya1972}) tells us that we may iterate $G$, starting from $\boldsymbol{x}=(0,\dots,0)$, and it will converge to a fixed point $\boldsymbol{q}\in[0,1]^{\hat L}$, where $q_\alpha$ is the extinction probability of the branching process if it starts with a single particle of type $\alpha$.
We now obtain the single-seed cascade probability as
\begin{equation}\label{eq-cascadeprob-final}
    \ptrig = 1 - \sum_{J,K\in\mathbb{N}^L} P(J,K) \prod_{\alpha\in\nicelayers} \big( (1-v_\alpha) +v_\alpha q_\alpha \big) ^{K_\alpha}.
\end{equation}

Instead of a single seed node, we can also consider a larger fixed number $n\in\mathbb{N}$ of seed nodes \cite{Gleeson2008a}.
In this case, the probability to trigger a macroscopic cascade is simply $1-(1-\ptrig)^n$, i.e.~the probability that any of the $n$ seed nodes succeed in triggering a cascade.

\subsection{Cascade Condition}
In addition to the cascade condition seen in section \ref{sec-size-cond}, one may also derive a cascade condition from the cascade probability \cite{Yagan2012}.
These two derivations yield different expressions, and we believe it is insightful to have both of them available for interpretation.

We derive the cascade condition by identifying the parameter regions where the single-seed cascade probability $\ptrig$ is positive.
According to the same theorem on multi-type branching processes we already used above, this is the case if and only if the leading eigenvalue of the previously mentioned mean matrix $M\in\mathbb{R}^{\hat L\times \hat L}$ is strictly greater than $1$.
Recall that the entries $M_{\alpha\beta}$ are defined as the expected number of type-$\beta$ offspring from a type-$\alpha$ particle in one generation.
This is related to the offspring generating function \eqref{eq-cascadeprob-offspring} by
\begin{equation}
    M_{\alpha\beta} = \frac{\textnormal{d}G_\alpha}{\textnormal{d}x_\beta}(1,\dots,1).
\end{equation}
We have
\begin{equation}
    \frac{\textnormal{d}G_\alpha}{\textnormal{d}x_\beta} =
    \frac{1}{v_\alpha z_\alpha} \sum_{\overset{J,K}{J_\alpha=1}} \hat P(J,K) \cdot K_\beta v_\beta((1-v_\beta) + v_\beta x_\beta)^{K_\beta-1} \cdot \prod_{\gamma\ne\beta} \big( (1-v_\gamma) +v_\gamma x_\gamma \big)^{K_\gamma},
\end{equation}
which for $x\to (1,\dots,1)$ converges to
\begin{equation}\label{eq-condition-matrix-micro}
     M_{\alpha\beta} = \frac{1}{v_\alpha z_\alpha} \sum_{\overset{J,K}{J_\alpha=1}} \hat P(J,K) \cdot K_\beta v_\beta,
\end{equation}
which is the expected number of vulnerable type-$\beta$-links that start at the endpoint of a randomly chosen vulnerable type-$\alpha$-link.
Thus, one can also confirm the possibility of single-seed cascades by checking that the dominant eigenvalue of $M$ exceeds $1$, instead of using $A$ from section \ref{sec-size-cond}.

\subsection{Simple Special Case}
\label{sec-prob-iid}
We will now once again consider the special case of i.i.d.~degrees, i.e.~we assume
\begin{equation}
    P(J,K) = \prod_{\alpha\in \layers} \pin(J_\alpha) \pout(K_\alpha).
\end{equation}
We further assume that there exist vulnerable links, i.e.~$\pin(1)>0$.
In this case $\hat L = L$, all layers have the same mean degree $z$, and the $v_\alpha$ all take the same value $v$, given by
\begin{equation}
    v = \frac{1}{z} \sum_{\overset{J}{J_\alpha=1}} \prod_\beta \pin(J_\beta) = \frac{1}{z} \pin(1).
\end{equation}
Additionally, the offspring generating functions \eqref{eq-cascadeprob-offspring} become independent of $\alpha$ (see Appendix \ref{apx-prob-iid}), allowing us to simplify the situation to a single generating function $G:\mathbb{R}\to\mathbb{R}$ that we iterate to arrive at a fixed point $q\in\mathbb{R}$.
This $G$ is given by
\begin{equation}
    G(x) = \gout((1-v)+vx)^{L},
\end{equation}
where $\gout$ is once again the probability generating function of the out-degree distribution $\pout$.
With the fixed point $q$, we obtain the cascade probability as
\begin{equation}
    \ptrig = 1 - \gout(1-v+vq)^{L} = 1-q.
\end{equation}
We can make some observations.
Firstly, both $\pin$ and $\pout$ affect the cascade probability, unlike the cascade size in section \ref{sec-size-iid}.
Second, while the entire distribution of $\pout$ appears in $\gout$, this is not the case for $\pin$:
Only $\pin(1)$ and the expected value $z$ are relevant.
So the tail behavior of the in-degree distribution does not affect the cascade probability, for the same reason as the cascade condition in section \ref{sec-size-iid}.
Furthermore, note that increasing either $L$ or $v$ will lead to an increase of $\ptrig$.
This corresponds to the intuitive understanding that adding more network layers opens more pathways for the spreading of cascades, as does a larger fraction of vulnerable links.

Moving on to the cascade condition, we find that the $M_{\alpha\beta}$ become
equal to the $A_{\alpha\beta}$ in section \ref{sec-size-iid} and thus $M_{\alpha\beta}=\pin(1)$ for all $\alpha,\beta\in\layers$.
Therefore, the cascade condition is once again
\begin{equation}
    L \cdot \pin(1) > 1.
\end{equation}
This is no coincidence: The two cascade conditions derived in sections \ref{sec-size-iid} and \ref{sec-prob-iid} are equivalent in general. See Appendix \ref{apx-proof-equivalence} for a proof.


\subsection{More Intuitive Cascade Condition}
Both versions of the cascade condition compare the leading eigenvalue $\lambda_A$ (resp.~$\lambda_M$) of a matrix $A$ (resp.~$M$) to $1$.
And while the individual entries of $A$ and $M$ can be directly interpreted in terms of the network, the same cannot be said easily for the eigenvalues.
So although both versions of the cascade condition are easy to compute, they do not add to our intuition about the interplay of network structure and cascades.
In this section, we provide a more interpretable -- but strictly weaker -- cascade condition.

We use the following consequence of the Perron-Frobenius theorem: for a non-negative matrix, the leading eigenvalue is bounded above (resp.~below) by the largest (resp.~smallest) row sum.
Denoting the row sums of $A$ by $r_\alpha(A)\defeq \sum_\beta A_{\alpha\beta}$, this means we have
\begin{equation}
    \min_\alpha r_\alpha(A) \le \lambda_A \le \max_\alpha r_\alpha(A),
\end{equation}
and also the analogous statement for $M$.
Note that $A_{\alpha\beta}$ is the probability that a randomly chosen type-$\alpha$-link depends directly on a vulnerable $\beta$-link.
We can thus interpret $r_\alpha(A)$ as the average number of vulnerable links on which a randomly chosen layer-$\alpha$-link directly depends.
Therefore, if the links of each type depend -- on average -- on more than one vulnerable link, the cascade condition is fulfilled.
Conversely, if the links of each type depend on less than one vulnerable link on average, the cascade condition is not fulfilled.
Note that this formulation is weaker than the original cascade condition, since it leaves open the case where $r_\alpha(A)$ is greater than one for some link types, while for others it is less.

We also obtain the analogous weak cascade condition for the row sums $r_\alpha(M)$ of $M$.
These may be interpreted as the average number of vulnerable links that directly depend on a randomly chosen vulnerable layer-$\alpha$-link.

Taken together, this gives us an intuition for some network structures that facilitate the growth of large cascades from microscopic seeds:
Such cascades will appear in networks where links typically have more than one vulnerable dependency, or where the vulnerable links typically lead to more than one other vulnerable link.
Conversely, we will not see cascades if links typically have less than one vulnerable dependency, or if vulnerable links typically lead to less than one other vulnerable link.

\subsection{In Constrained Multiplex Networks}
\label{sec-prob-constrained}
We briefly revisit the assumptions that were needed to avoid degeneracy of the branching process, and we adapt them to the special case of constrained multiplex networks.
Non-singularity is fairly straightforward:
For a constrained multiplex network to result in a singular branching process, it has to satisfy $\pin(1)=\pout(1)=1$.
To ensure agreement of the mean in- and out-degrees, for every layer $\alpha\in\layers$ there needs to exist exactly one $\beta$ such that $C_{\beta\alpha}=1$, while all other entries of $C$ are $0$.
Thus, by demanding non-singularity we only exclude highly degenerate networks where every node has exactly one in- and one out-neighbor.
We turn to positive regularity, which now becomes more concrete than in the general case.
Recall that positive regularity is a statement about the mean matrix $M\in\mathbb{R}^{L\times L}$, whose entry $M_{\alpha\beta}$ is given by the expected number of type-$\beta$ vulnerable links starting at the endpoint of a type-$\alpha$ vulnerable link.
The process is positive regular if there exists $n\in\mathbb{N}$ s.t.~all entries of $M^n$ are strictly positive.
Non-negative matrices with this property are sometimes referred to as \textit{primitive}~\cite{Seneta1981}.
Whether a non-negative matrix is primitive only depends on the positions of its positive entries, not on the values of those entries.
This is helpful in our case, since (assuming $\pin(1)>0$) it is easy to see that $M_{\alpha\beta}>0$ if and only if $C_{\alpha\beta}>0$.
Positive regularity of the branching process is therefore equivalent to primitivity of the constraint matrix.
One can even show that primitivity of the constraint matrix implies non-singularity.

We now assume that $C$ is primitive and determine the single-seed cascade probability. We only state the results, the calculations can be found in Appendix \ref{apx-constrained-derivations}.
Substituting \eqref{eq-p-constrained} in to \eqref{eq-cascadeprob-v} yields
\begin{equation}
    v_\alpha = \frac{\pin(1)}{z}.
\end{equation}
Since this is now independent of $\alpha$, we will drop the subscript and refer to $v\defeq\frac{\pin(1)}{z}$ from now on.
For the offspring-generating function \eqref{eq-cascadeprob-offspring} we obtain
\begin{equation}
G_\alpha(\mathbf{x})
    = \left( \sum_\beta C_{\alpha\beta} \right)^{-1} \sum_\gamma C_{\alpha\gamma}\cdot \gout^{Lz_\gamma}\big( (1-v)+vx_\gamma \big),
\end{equation}
where $\gout^z$ is the PGF of $\pout^z$.
Finally, the cascade probability \eqref{eq-cascadeprob-final} becomes
\begin{equation}
    \ptrig = 1-\frac{1}{L} \sum_\alpha \gout^{Lz_\alpha} \big( (1-v) + vq_\alpha \big),
\end{equation}
where $\mathbf{q}\in[0,1]^L$ is the fixed point of the combined $G$ as before.


\section{Novel Phenomena}
\label{sec-constraint-examples}
Typically, the Watts threshold model and its multiplex variants undergo two phase transitions upon variation of the link density of the network \cite{Watts2002, Brummitt2012, Yagan2012}.
At a relatively low link density, there is a continuous phase transition of the cascade size as the network becomes connected, analogous to percolation.
At a higher density, cascades disappear in a first-order phase transition as the increasing degree of nodes makes them too robust to propagate cascades.

In this section, we will see some of the impacts the constrained multiplex structure can have on this picture.
The first of these examples can even be understood analytically via a center manifold reduction, while the others will only be investigated numerically.

Note that the effects presented here truly require the multiplex formulation, i.e.~they cannot be reproduced by a monoplex Watts model with similar parameters (see Appendix \ref{apx-projected-theory}).

\subsection{Explosive Onset of Cascades}
In our constrained multiplex networks, we find that the continuous transition where cascades emerge may become first-order, i.e.~cascades of a significant size may appear suddenly upon variation of the network density.
Such an explosive onset of cascades has been reported before, facilitated by changes to the dynamics of the model, such as heterogeneity in the thresholds \cite{Gleeson2007} or in the response rules of nodes \cite{Lee2014}. Indeed, such a change from a nonexplosive to an explosive transition is in a sense expected when modifying a model \cite{Kuehn2021}.
In contrast to the mentioned prior works, we do not alter the node dynamics, but rather the structure of the underlying network.
In doing so, we add to previous observations about the multitude of ways in which specific multiplex network structures may change the phase transitions, e.g.~by interrupting the cascade region \cite{Zhuang2017} or by creating additional cascade regions \cite{Unicomb2019}.

Our example is constructed as follows.
We let $\pin = \Pois(z)$ be a Poisson distribution with mean $z$, which will be our control parameter.
Likewise, we let the $(\pout^{\hat z})_{\hat z>0} =(\Pois(\hat z))_{\hat z>0}$.
The constraint matrix depends on a parameter $p\in[0,1]$, and is given by
\begin{equation}
    C_p\defeq\pmat{1&1&p\\1&1&p\\1&1&p}.
\end{equation}
The parameter $p$ will allow us to vary the dominant eigenvalue of $C_p$, thus controlling whether the cascade condition \eqref{eq-condition-constrained} is fulfilled.
We will find the first-order transition when the value of $p$ is in a range near this decision boundary.
To ensure that condition \eqref{eq-condition-constrained} is exact, we consider $\rho_0 = 0$.

As the dominant eigenvalue of $C_p$ is given by $2+p$, condition \eqref{eq-condition-constrained} will be fulfilled whenever $\pin(1)=z e^{-z} > (2+p)^{-1}$.
Since $ze^{-z}$ takes its maximum value $e^{-1}$ at $z=1$, the condition can only be fulfilled for $p\in(e-2,1]$.
For any such $p$, there are two values of $z$ that solve the equation $ze^{-z}=(2+p)^{-1}$, corresponding to the two phase transitions marking the emergence and disappearance of cascades.
We are interested in the smaller of these solutions, which we will call $z^\star(p)$ to make its dependence on $p$ explicit.

We will use bifurcation theory to determine whether this transition is continuous or discontinuous.
For this, we interpret equation \eqref{eq-cascadesize-edge-constrained} as a parameter-dependent iterated map $x\mapsto F(x,z)$, where $x\in\mathbb{R}^L,z\in[0,\infty)$ and the components of $F$ are given by
\begin{equation}
    F_\alpha(x,z)
    = 1 - (1-\rho_0)\prod_{\beta\in\layers} 
        1+C_{\beta\alpha}\cdot\Big(\gin(0)-\gin(x_\beta)\Big).
\end{equation}
Since $\pin$ is a Poisson distribution with mean $z$, we have $\gin(s)=e^{z(s-1)}$, which is where the dependence on $z$ comes in.

The bifurcation we are interested in is located at $(x,z)=(\mathbf{0}, z^\star(p))$.
We will show that this is a transcritical bifurcation and determine its normal form coefficients.
From these coefficients we can then read off for which values of $p$ the transition is explosive.

First, it is clear that $F(\mathbf{0},z)=\mathbf{0}$ and thus $\mathbf{0}$ is a fixed point for all $z\in[0,\infty)$.
Second, we already know from the derivation of the cascade condition that the Jacobian of this fixed point is $A = C_p\cdot\pin(1)$,
and we have defined $z^\star$ such that the dominant eigenvalue of $A$ is $1$ and the fixed point is non-hyperbolic.
The other two eigenvalues are $0$, so the center eigenspace of this bifurcation is $1$-dimensional.
This center eigenspace is spanned by the eigenvector $q=\pmat{1&1&1}^\top$.
To ensure that the dominant eigenvalue passes through $1$ transversally, we compute $\frac{\mathrm{d}}{\mathrm{d}z} A = C_p\cdot\frac{\mathrm{d}}{\mathrm{d}z}\pin(1)=C_p\cdot(1-z)e^{-z}$.
The dominant eigenvalue of this matrix is the derivative of the dominant eigenvalue of $A$ w.r.t.~$z$, and it is given by $(2+p)\cdot(1-z^\star(p))e^{-z^\star(p)} \eqdef c_{xz}$, which is strictly positive for all $p\in(e-2,1]$.

Since the center eigenspace of the bifurcation has dimension $1$, we make the ansatz that the center manifold can be parameterized by
\begin{equation}
    H(\xi) = \xi q + \frac{1}{2} \xi^2 h_2 + O(\xi^3),
\end{equation}
where $h_2\in\mathbb{R}^3$ is a vector whose explicit value we will not require.
So far, we have established that the restriction of $F$ to the center manifold takes the form
\begin{equation}
    \xi\ \mapsto\ \xi + c_{xz}\xi \big(z-z^\star(p)\big) + c_{xx}\xi^2 + \mathrm{h.o.t.}\ ,
\end{equation}
where the coefficient $c_{xx}$ is still unknown and the higher-order terms have total degree at least $3$ in $\xi,z$.
To decide whether the transition is explosive or not, we need to determine the sign of $c_{xx}$.
To this end, we use a combined reduction/normalization technique that avoids explicitly computing a Taylor expansion of the center manifold \cite{Kuznetsov2023}.
The calculation can be found in Appendix \ref{apx-explosive-cxx}, and we find that
\begin{equation}
    c_{xx} = \frac{1}{(2+p)^2}\cdot\left(\frac{1}{2} \left(2+p^2\right) \cdot z^\star(p) - \frac{2+p^3}{2+p}\right)
\end{equation}
Note that $0<c_{xx} \iff 1<\frac{2+p}{2}\cdot\frac{2+p^2}{2+p^3}\cdot z^\star(p) \eqdef f(p)$, and we easily see that $f(p)>1$ for $p\to e-2$.
Thus, for $p$ near $e-2$ we have $c_{xz}>0$ and $c_{xx}>0$, meaning that our transcritical bifurcation corresponds to an explosive transition.
This occurs close to the regime $p<e-2$ where cascades are not possible for any value of $z$.
For completeness, we note that the transition becomes continuous for larger values of $p$, as it can be shown that $f(1)<1$ and thus $c_{xx}<0$ for $p=1$.

\subsection{Nested Cascade Regions}
\label{sec-example-nested}
We now present an example of a simple constrained multiplex network structure that exhibits an additional phase where especially large cascades are possible.
Like in the preceding section, we let the in- and out-degree distributions be Poisson, and we choose the mean $z$ of $\pin$ as our control parameter.
Let $p_1,p_2\in[0,1]$ and consider the constraint matrix
\begin{equation}
C = \pmat{
    p_1 & p_1 & p_1 & 0   & 0   & 0  \\
    p_1 & p_1 & p_1 & 0   & 0   & 0  \\
    p_1 & p_1 & p_1 & 0   & 0   & 0  \\
    0   & 0   & p_2 & p_2 & p_2 & p_2\\
    0   & 0   & p_2 & p_2 & p_2 & p_2\\
    0   & 0   & p_2 & p_2 & p_2 & p_2\\
}.
\end{equation}
This corresponds to a network where there are two groups of node types, and
nodes may have incoming links from every type of their group with probability $p_1$ or $p_2$ respectively.
There is one type of nodes in the first group which may additionally have links from one type in the second group.
The resulting networks are only weakly connected, with two distinct giant strongly connected components corresponding to the two groups.
Notice that the cascade can only propagate from the second component to the first, but not vice versa.
This corresponds to the fact that $C$ is not primitive and therefore, the assumption of positive regularity in our computation of the cascade probability is violated.
One could, of course, also consider variants of this example with more groups, or groups containing different numbers of node types.
Here we have deliberately chosen a small example to reduce the computational cost of simulations.

For our first series of experiments, we choose the parameters to be $p_1=1.0$ and $p_2=0.91$.
This means that nodes in the second component may have fewer dependencies, making the second component less vulnerable to cascades than the first.
However, when a cascade does spread in the second component, it can reach the entire network, while cascades starting in the first component stay confined to that component.

\begin{figure}
    \centering
    \begin{overpic}[width=0.5\linewidth]{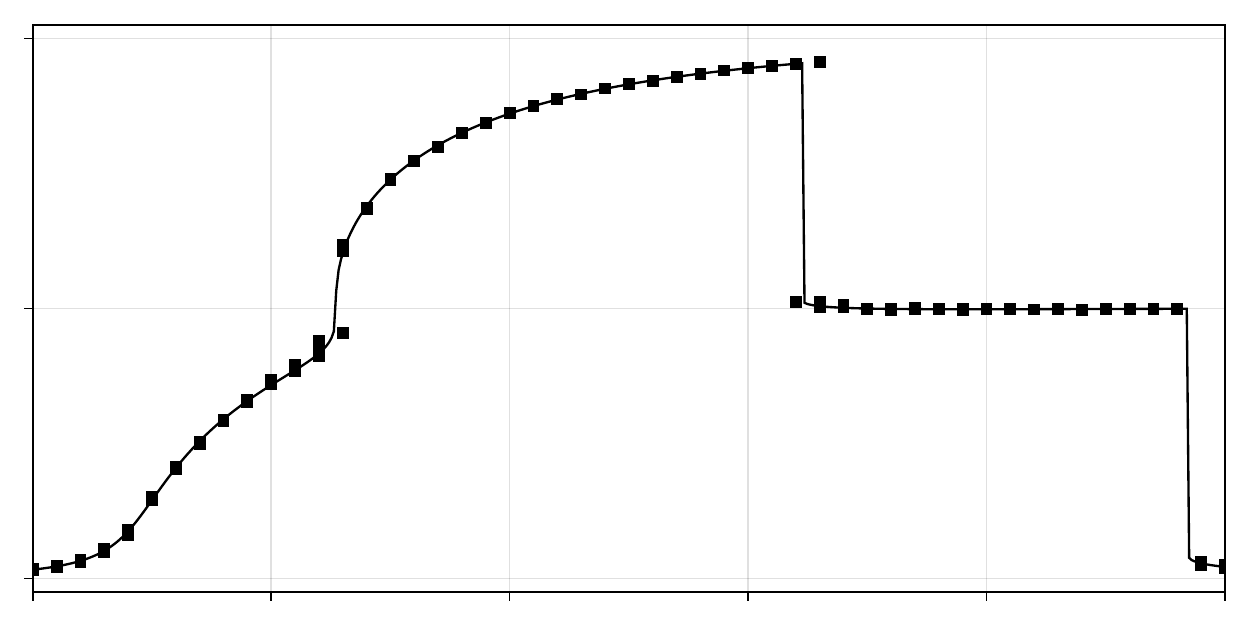}
        \put(-4,3){\ticklabelsize $0.0$}
        \put(-4,24.5){\ticklabelsize $0.5$}
        \put(-4,46){\ticklabelsize $1.0$}
        \put(-1,-1){\ticklabelsize $0.50$}
        \put(18,-1){\ticklabelsize $0.75$}
        \put(37,-1){\ticklabelsize $1.00$}
        \put(56,-1){\ticklabelsize $1.25$}
        \put(75,-1){\ticklabelsize $1.50$}
        \put(94,-1){\ticklabelsize $1.75$}
        \put(50,-4.5){$z$}
        \put(-9,10){\rotatebox{90}{cascade size}}
    \end{overpic}
    \caption{Sizes of macroscopic cascades for $p_1=1.0,\ p_2=0.91$, $\rho_0=10^{-3}$ and various values of $z$.
            The solid line is the cascade size from the analysis, the squares show the cascade sizes observed in our simulations.}
    \label{fig:nested-a-macro}
\end{figure}

We first investigate the case of macroscopic seeds.
We have used equations \eqref{eq-cascadesize-node-constrained} and \eqref{eq-cascadesize-edge-constrained} to compute the expected size of cascades starting from an initial seed size of $\rho_0=10^{-3}$, for different values of $z$.
For comparison, we also performed simulations on networks of $5\cdot10^6$ nodes.
One network was generated for each value of $z$, then $10$ cascades with different seeds were simulated on that network.
The results are depicted in Figure \ref{fig:nested-a-macro}, showing good agreement between simulations and theory.
Note that the figure shows all $10$ outcomes for each $z$, but they are usually too close to tell them apart.
We can see that there is an additional phase of especially large cascades nested inside a broader region of smaller cascades.
This is easily explained, as
the smaller phase is the range of values of $z$ where the second component allows for cascades to spread.
Cascades in this regime will encompass both components and thus reach almost the entire network, while cascades for other values of $z$ are limited to the first component which takes up only half the network.

\begin{figure}
    \centering
    \subfloat{
    \begin{overpic}[width=0.595\linewidth]{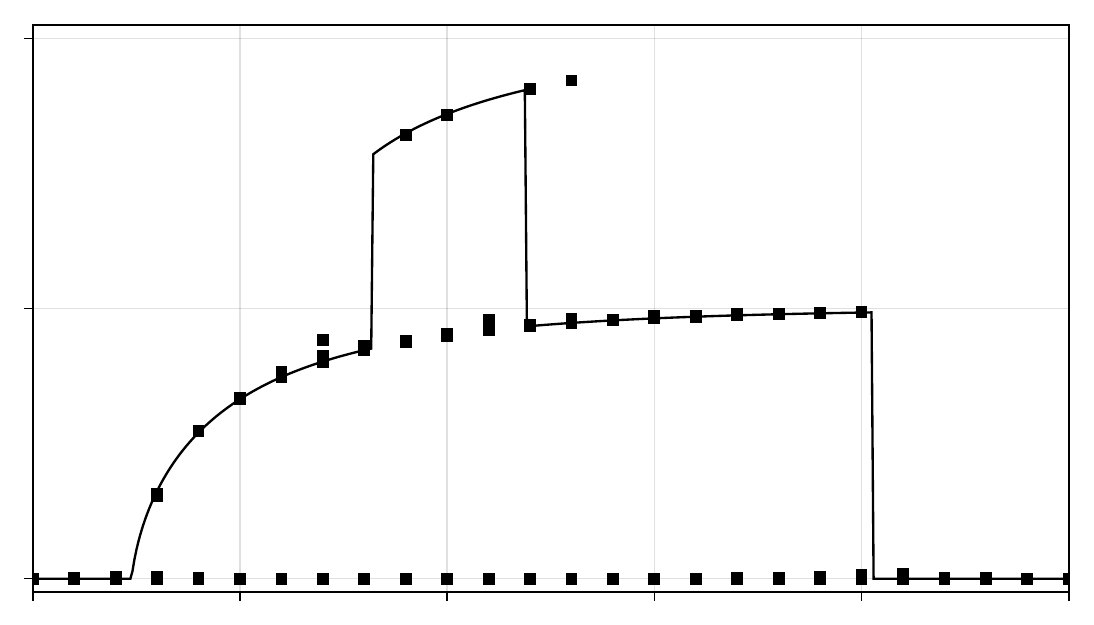}
        \put(-4,3){\ticklabelsize $0.0$}
        \put(-4,28){\ticklabelsize $0.5$}
        \put(-4,52){\ticklabelsize $1.0$}
        \put(-1,-1){\ticklabelsize $0.50$}
        \put(18,-1){\ticklabelsize $0.75$}
        \put(37,-1){\ticklabelsize $1.00$}
        \put(56,-1){\ticklabelsize $1.25$}
        \put(75,-1){\ticklabelsize $1.50$}
        \put(94,-1){\ticklabelsize $1.75$}
        \put(50,-4){$z$}
        \put(-9,15){\rotatebox{90}{cascade size}}
        \put(5,46){(a)}
    \end{overpic}}\quad
    \subfloat{
    \begin{overpic}[width=0.34\linewidth]{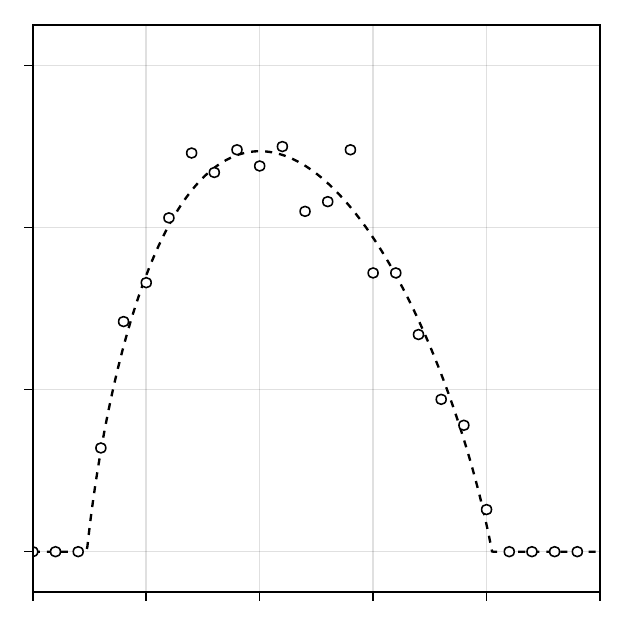}
        \put(-5,10){\ticklabelsize $0.0$}
        \put(-5,36){\ticklabelsize $0.1$}
        \put(-5,62){\ticklabelsize $0.2$}
        \put(-5,88){\ticklabelsize $0.3$}
        \put(-1,-1.5){\ticklabelsize $0.50$}
        \put(18,-1.5){\ticklabelsize $0.75$}
        \put(37,-1.5){\ticklabelsize $1.00$}
        \put(54,-1.5){\ticklabelsize $1.25$}
        \put(72,-1.5){\ticklabelsize $1.50$}
        \put(90,-1.5){\ticklabelsize $1.75$}
        \put(50,-8){$z$}
        \put(-13,14){\rotatebox{90}{cascade probability}}
        \put(8,81){(b)}
    \end{overpic}}
    \vspace{0.1em}
    \caption{Single-seed cascades for $p_1=1.0,\ p_2=0.91$ and various values of $z$.
    (a) Sizes.
            Solid line is the cascade size from the macroscopic analysis, squares show the sizes of cascades observed in simulations.
    (b) Probability.
            The dashed line shows $\ptrig$ from the microscopic analysis, the circles show the fraction of observed cascades that reached more than $1\%$ of the network.}
    \label{fig:nested-a-micro}
\end{figure}

We now turn our attention to the case of microscopic seeds and encounter a subtlety:
It is generally assumed -- and often observed -- that cascades spreading from microscopic seeds either stay vanishingly small, or reach the size that is predicted by the macroscopic analysis in the limit $\rho_0\to 0$.
But in this example, the maximum size of a cascade depends on the component in which we choose the seed node.
While cascades starting from a seed node in the second component may reach the full size predicted by the macroscopic theory, those which start in the first component remain confined to the first component.
To demonstrate this, we have conducted simulations where a network with $10^6$ nodes was generated for different values of $z$, then single-seed cascades were computed from $500$ different seed nodes in each network.
In Figure \ref{fig:nested-a-micro}a, we can clearly see the two distinct sizes of global cascades, of which only the larger one corresponds to the macroscopic theory.
Furthermore, this larger cascade size is also observed much less frequently than the smaller one:
For $z=1.0$, less than $1\%$ of seed nodes led to cascades of the largest size, while around $25\%$ reached the intermediate size.
The reason for this discrepancy is that there are effectively two different single-seed cascade probabilities at play.
One for seed nodes in the first component, and one for seed nodes in the second.
Despite this complication, we find that the analytically computed $\ptrig$ agrees well with simulations (see Fig.~\ref{fig:nested-a-micro}b).
This demonstrates that the microscopic analysis can work well even in cases where the assumptions presented in section \ref{sec-prob} are violated, although it remains unclear just how far its validity extends.

\begin{figure}
    \centering
    \begin{overpic}[width=0.5\linewidth]{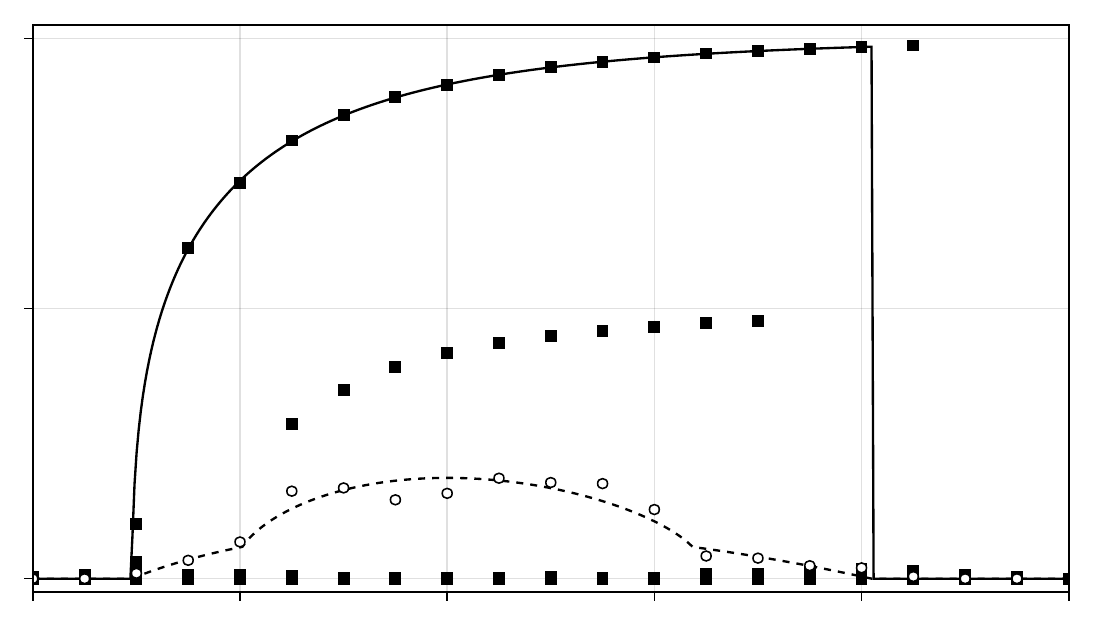}
        \put(-4,3){\ticklabelsize $0.0$}
        \put(-4,28){\ticklabelsize $0.5$}
        \put(-4,52){\ticklabelsize $1.0$}
        \put(-1,-1){\ticklabelsize $0.50$}
        \put(18,-1){\ticklabelsize $0.75$}
        \put(37,-1){\ticklabelsize $1.00$}
        \put(56,-1){\ticklabelsize $1.25$}
        \put(75,-1){\ticklabelsize $1.50$}
        \put(94,-1){\ticklabelsize $1.75$}
        \put(50,-4.5){$z$}
        \put(-10,6){\rotatebox{90}{size \& probability}}
    \end{overpic}
    \caption{Probability and sizes of single-seed cascades for $p_1=0.94,\ p_2=1.0$ and various values of $z$.
            Solid line is the cascade size from the macroscopic analysis, dashed line is $\ptrig$ from the microscopic analysis.
            Squares show the sizes of cascades observed in simulations. 
            Circles show the fraction of observed cascades that reached more than $1\%$ of the network.
            }
    \label{fig:nested-b}
\end{figure}

The behavior observed in this example results from the fact that cascades may only spread from the more robust second component to the more fragile first component.
We can also consider the reverse case where the first component is more robust, while the second one is fragile.
For this, we choose the parameters $p_1=0.94$ and $p_2=1.0$.
We examine the microscopic case using the same procedure as before, generating one network of $10^6$ nodes for each value of $z$, then simulating cascades from $500$ different seed nodes.
As we can see in Figure \ref{fig:nested-b}, the single-seed cascades in this network also have two different sizes, and the macroscopic analysis again only reveals the larger one.
We also see a noticeable increase in the cascade probability in the phase where the intermediate cascade size is possible.
This reflects the more balanced probabilities of both cascade sizes, as both the large and the intermediate size were reached from about $10\%$ of seed nodes at $z=1.0$.

Although the network structure underlying these examples is very simple, the following two conclusions apply more broadly.
First, networks with multiple giant strongly connected components will exhibit multiple single-seed cascade sizes, and the typical analytic approach for determining the cascade size will be of limited applicability.
Indeed, one can even construct networks where none of the single-seed cascade sizes match the analytic prediction from the macroscopic theory: Just think of three components linked in a
\mbox{$\circ\rightarrow\circ\leftarrow\circ$} pattern.
Second, if a network has a structure containing different \enquote{parts} with different susceptibility to cascades, we expect to see this heterogeneity reflected in additional phase transitions of the cascade size or probability.

\subsection{Central Cusp}
We now briefly give an example of a network with a unique giant strongly connected component that nonetheless contains parts with different robustness.
This leads to a similar \enquote{nested} phase of increased cascade size as in the previous section, but without the appearance of multiple different single-seed cascade sizes.

Let the in- and out-degree distributions again be Poisson, and choose the mean $z$ of $\pin$ as our control parameter.
Consider the constraint matrix
\begin{equation}
C = 
\left(\begin{array}{ccc|c|ccc}
  &&&q_2          &
\\&Q_1&&\vdots    &&0&
\\&&&q_2          &
\\\hline
  q_1&\dots&q_1&q_2&q_2&\dots&q_2
\\\hline
  &&&q_2          
\\&0&&\vdots    &&Q_2
\\&&&q_2          
\end{array}\right),
\end{equation}
where $q_1,q_2\in[0,1]$ are parameters, $Q_1$ is the $20\times 20$ matrix whose every entry is $q_1$, and $Q_2$ is the $20\times 20$ matrix with every entry $q_2$.

The idea here is similar to the construction of the previous example.
Again, there are two groups of node types such that nodes may depend on all types in the same group with a certain probability.
But this time, there is an additional node type which is considered part of both groups.
The resulting networks have a unique giant strongly connected component, and cascades are able to propagate in either direction between the two groups.
This network structure yields interesting behavior for various choices of the parameters, and we have chosen $q_1=0.95,\ q_2=0.131$ to show here.

\begin{figure}
    \centering
    \begin{overpic}[width=0.5\linewidth]{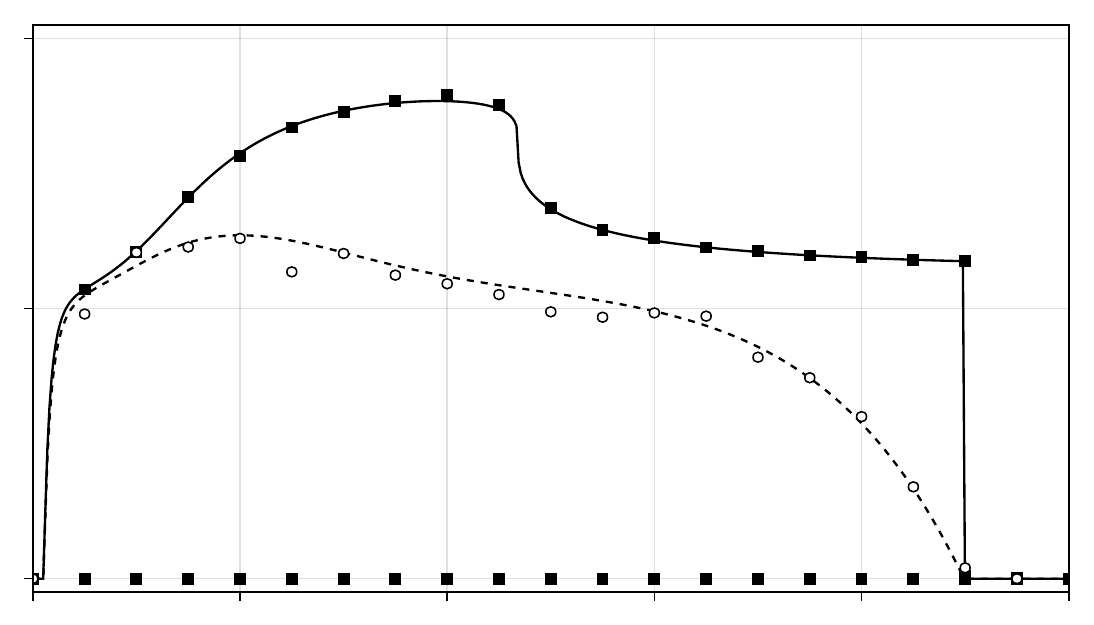}
        \put(-4,3){\ticklabelsize $0.0$}
        \put(-4,28){\ticklabelsize $0.5$}
        \put(-4,52){\ticklabelsize $1.0$}
        \put(2,-1){\ticklabelsize $0$}
        \put(21,-1){\ticklabelsize $1$}
        \put(40,-1){\ticklabelsize $2$}
        \put(59,-1){\ticklabelsize $3$}
        \put(77.5,-1){\ticklabelsize $4$}
        \put(97,-1){\ticklabelsize $5$}
        \put(50,-4.5){$z$}
        \put(-10,6){\rotatebox{90}{size \& probability}}
    \end{overpic}
    \caption{Probability and sizes of single-seed cascades for $q_1=0.95,\ q_2=0.131$ and various values of $z$.
            Solid line is the cascade size from the macroscopic analysis, dashed line is $\ptrig$.
            Squares show the sizes of cascades observed in simulations,
            circles show the fraction of observed cascades that reached more than $1\%$ of the network.}
    \label{fig:twoish-micro}
\end{figure}

We once again simulated single-seed cascades and compared the observed cascade sizes and probabilities to those obtained from the analysis.
The simulations were done on networks with $10^5$ nodes, and from $500$ different seed nodes on each network.
Simulations with macroscopic seeds were also performed, but did not yield anything of further interest.
Results are shown in Figure \ref{fig:twoish-micro}.

\begin{figure}
    \centering
    \begin{overpic}[width=0.5\linewidth]{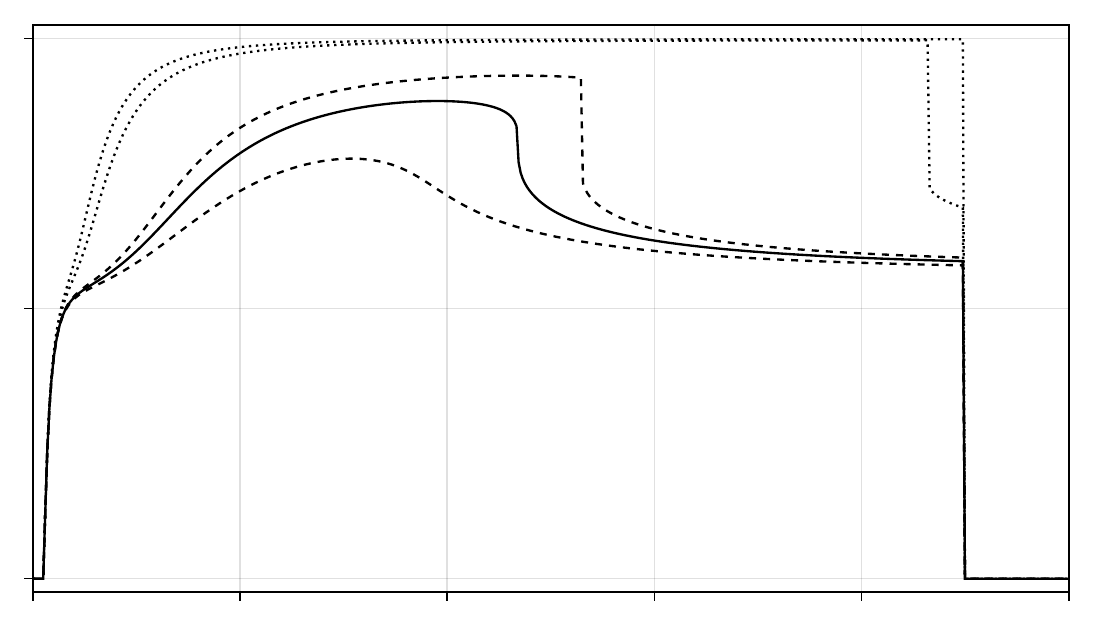}
        \put(-4,3){\ticklabelsize $0.0$}
        \put(-4,28){\ticklabelsize $0.5$}
        \put(-4,52){\ticklabelsize $1.0$}
        \put(2,-1){\ticklabelsize $0$}
        \put(21,-1){\ticklabelsize $1$}
        \put(40,-1){\ticklabelsize $2$}
        \put(59,-1){\ticklabelsize $3$}
        \put(77.5,-1){\ticklabelsize $4$}
        \put(97,-1){\ticklabelsize $5$}
        \put(50,-4.5){$z$}
        \put(28,39){\tiny $0.12$}
        \put(35,45){\tiny $0.131$}
        \put(54,45){\tiny $0.14$}
        \put(77,45){\tiny $0.22$}
        \put(89,45){\tiny $0.25$}
        \put(-10,13){\rotatebox{90}{cascade size}}
    \end{overpic}
    \caption{Analytic cascade size for $q_1=0.95$ and different values of $q_2$.
        The solid line shows $q_2=0.131$, the upper and lower dashed lines show $q_2=0.14$ and $q_2=0.12$, respectively.
        The upper and lower dotted lines show $q_2=0.25$ and $q_2=0.22$, respectively.}
    \label{fig:twoish-cusp}
\end{figure}

We see that the cascade region has been split by a new phase transition,
with large global cascades to its left and somewhat smaller cascades on its right.
Intuitively, the larger cascades again occur because they spread in both the vulnerable first group of node types as well as the more robust second group.
If both groups are equally vulnerable to cascades (e.g.~$q_1=q_2=0.95$), we do not see the transition.
But if we then decrease $q_2$, the transition initially appears as a first-order transition before disappearing via a cusp (see Fig.~\ref{fig:twoish-cusp}).



\section{Conclusion}
We have presented a comprehensive analysis of the or-rule threshold model for threshold $\phi=1$ on directed multiplex configuration model networks, deriving the expected cascade size, single-seed cascade probability and cascade condition.
To isolate the effects of functional structure, we introduced the Constrained Multiplex Network Model (CMNM), which is parameterized to separate the degree distribution from the functional structure as much as possible.
We found that even when the marginal degree distributions are all Poisson, different choices of functional structure can induce drastically different phase transitions in the model.

This should be relevant to anyone who wishes to model cascade-like processes in real-world systems that feature complementarity:
It is vitally important that such models account for the \textit{correct} functional structure underlying the real system.
Models similar to the CMNM may be a convenient tool for this, depending on the use case.

Of course, the findings presented here do not yet constitute a complete picture of functional structure, and they must be complemented by further research.
Most importantly, one should examine the impact of complementarity on other dynamical processes besides threshold models,
such as simple epidemic contagion, synchronization, interdependent networks, or overload cascades.
In addition, the CMNM may not be the best model for networks with functional structure, as it makes some simplifying assumptions.
It assumes that each node only has out-links on a single layer, i.e.~it only fulfills a single function.
It also assumes that the marginal degree distributions of each node on each layer are the same, and in our experiments we fixed all these distributions to be exclusively Poisson.
While these assumptions are mathematically convenient, they are far too restrictive for a general model.
Lastly, we also uncovered a more technical question about the analysis of threshold models:
In section \ref{sec-example-nested} we have seen that it is not always possible to determine the size of single-seed cascades by considering the $\rho_0\to 0$ limit of the macroscopic theory.
Particularly in weakly connected networks, the cascades may reach several distinct sizes, meaning that we are really looking at a nontrivial distribution of global cascade sizes.
This indicates a need to develop a new method of calculating this distribution of single-seed cascade sizes.


\section*{Acknowledgments}
This work was supported by the DFG through the project \enquote{Stochastic Epidemic-Economic Adaptive Network Dynamics}, project number 496237661.
CKl wishes to thank the Complexity Science Hub Vienna for hospitality.


\printbibliography

\newpage
\appendix
\section{Derivations for i.i.d.~Degrees}
We will repeatedly use the following statement on products of series:
\begin{lemma}\label{lemma-sumprod}
    Let $\mathcal{I}_1,\dots,\mathcal{I}_k\subseteq\mathbb{N}$,
    and for $i\in\{1,\dots,k\}, j\in\mathcal{I}_i$ let $a_{i,j}\in\mathbb{R}$ such that for every $i$,
    the series $\sum_{j\in\mathcal{I}_i}a_{i,j}$ is absolutely convergent.
    Then one has
    \begin{equation}
        \prod_{i=1}^k \sum_{j\in\mathcal{I}_i} a_{i,j} \quad = \sum_{J\in\mathcal{I}_1\times\cdots\times\mathcal{I}_k} \prod_{i=1}^k a_{i,J_i}.
    \end{equation}
\end{lemma}

\subsection{Cascade Size}\label{apx-size-iid}
We substitute
\begin{equation}
    P(J,K)=\prod_{\alpha\in \layers} \pin(J_\alpha)\pout(K_\alpha)
\end{equation}
into
\begin{equation}
     q_\alpha(t+1)=1- \frac{1-\rho_0}{z_\alpha} \sum_{J,K} P(J,K) K_\alpha \prod_{\overset{\beta\in \layers}{J_\beta>0}}\left( 1- q_\beta(t)^{J_\beta} \right)
\end{equation}
and calculate
\begin{align}
    &1- \frac{1-\rho_0}{z_\alpha} \sum_{J,K\in\mathbb{N}^L} P(J,K) K_\alpha \prod_{\overset{\beta\in \layers}{J_\beta>0}}\left( 1- q_\beta(t)^{J_\beta} \right)
    \\={}&
    1-\frac{1-\rho_0}{z_\alpha} \cdot \sum_{J,K\in\mathbb{N}^L}\left( K_\alpha \left(\prod_{\gamma\in \layers} \pin(J_\gamma)\pout(K_\gamma) \right) \prod_{\overset{\beta\in \layers}{J_\beta>0}}\left( 1- q_\beta(t)^{J_\beta} \right) \right)
    \\={}&
    1-\frac{1-\rho_0}{z_\alpha} \cdot
        \sum_{K\in\mathbb{N}^L} K_\alpha \left(\prod_{\gamma\in \layers} \pout(K_\gamma) \right)
        \sum_{J\in\mathbb{N}^L} \left(\prod_{\gamma\in \layers} \pin(J_\gamma) \right)
        \prod_{\overset{\beta\in \layers}{J_\beta>0}}\left( 1- q_\beta(t)^{J_\beta} \right).
\end{align}
Applying Lemma \ref{lemma-sumprod} to the first sum and merging the last two product terms yields
\begin{align}
    ={}&
    1-\frac{1-\rho_0}{z_\alpha} \cdot \sum_{k\in\mathbb{N}}\Big( k \pout(k) \Big)
        \cdot \sum_{J\in\mathbb{N}^L} \left(\prod_{\beta\in \layers} \pin(J_\beta)
        \begin{cases}
        1-q_\beta(t)^{J_\beta} & \text{if } J_\beta > 0\\
        1                      & \text{if } J_\beta = 0
        \end{cases}
        \right).
\end{align}
We apply Lemma \ref{lemma-sumprod} again, this time to the other sum:
\begin{align}
    ={}&
    1-\frac{1-\rho_0}{z_\alpha} \cdot z_\alpha
        \cdot \prod_{\beta\in \layers} \left(\sum_{j\in\mathbb{N}} \pin(j)
        \begin{cases}
        1-q_\beta(t)^{j} & \text{if } j > 0\\
        1                & \text{if } j = 0
        \end{cases}
        \right)
    \\={}&
    1-(1-\rho_0) \cdot \prod_{\beta\in \layers}\left( \pin(0)+ \sum_{j>0} \pin(j)\cdot \left( 1-q_\beta(t)^j \right) \right)
    \\={}&
    1-(1-\rho_0) \cdot \prod_{\beta\in \layers}\left( 1 -\sum_{j>0} \pin(j)\cdot q_\beta(t)^j \right).
\end{align}
We notice that this no longer depends on $\alpha$, and we reduce the system to a single variable $q(t)\defeq q_\alpha(t) (\forall \alpha\in \layers)$. We obtain
\begin{equation}
    q(t+1) = 1-(1-\rho_0) \cdot \left( 1 -\sum_{j>0} \pin(j)\cdot q(t)^j \right)^{L},
\end{equation}
which can be written concisely in terms of the generating function $\gin$ of $\pin$. We finally arrive at
\begin{equation}
    q(t+1) = 1-(1-\rho_0)\cdot \Big( 1 + \gin(0) -\gin(q(t))  \Big) ^{L}.
\end{equation}
The calculation for $\rho(t+1)$ is analogous and yields exactly the same right-hand side.
We therefore have $\rho(t)=q(t)$ for all $t$ and there is no need for two separate equations.

\subsection{Cascade Condition from Size}\label{apx-cond-iid}
We substitute
\begin{equation}
    P(J,K)=\prod_{\alpha\in \layers} \pin(J_\alpha)\pout(K_\alpha)
\end{equation}
into
\begin{equation}
    A_{\alpha\beta} = \frac{1}{z_\alpha} \sum_{\overset{J,K}{J_\beta=1}} P(J,K) K_\alpha,
\end{equation}
and calculate
\begin{align}
A_{\alpha\beta}
    &= \frac{1}{z_\alpha} \cdot \sum_{\overset{J,K \in \mathbb{N}^L}{J_\beta=1}} \left( K_\alpha \prod_{\gamma\in \layers} \pin(J_\gamma) \pout(K_\gamma) \right)\label{eq-apx-cond-iid-1}\\
    &= \frac{1}{z_\alpha} \cdot \sum_{k\in\mathbb{N}}\Big( k \pout(k) \Big)
        \cdot \sum_{\overset{J\in\mathbb{N}^L}{J_\beta=1}} \left( \prod_{\gamma\in \layers} \pin(J_\gamma) \right)\\
    &= \frac{1}{z_\alpha} \cdot z_\alpha \cdot \pin(1)\\
    &= \pin(1).
\end{align}

\subsection{Cascade Probability}\label{apx-prob-iid}
We substitute
\begin{equation}
    P(J,K)=\prod_{\alpha\in \layers} \pin(J_\alpha)\pout(K_\alpha)
\end{equation}
into the offspring generating functions
\begin{equation}
    G_\alpha(\boldsymbol{x}) = \frac{1}{v z} \sum_{\overset{J,K\in\mathbb{N}^{L}}{J_\alpha=1}} P(J,K) \prod_{\beta\in\layers} \big( (1-v_\beta) +v_\beta x_\beta \big)^{K_\beta}.
\end{equation}
and find
\begin{align}
    G_\alpha(x)
        &= \frac{1}{vz} \sum_{\overset{J,K\in\mathbb{N}^L}{J_\alpha=1}} \prod_{\beta\in \layers} \pin(J_\beta)\pout(K_\beta) ((1-v)+vx_\beta)^{K_\beta}\\
        &= \frac{1}{vz} \sum_{\overset{J\in\mathbb{N}^L}{J_\alpha=1}} \prod_{\beta\in \layers} \pin(J_\beta) \cdot \sum_{K\in\mathbb{N}^L} \prod_{\beta\in \layers}\pout(K_\beta) ((1-v)+vx_\beta)^{K_\beta}\\
        &= 1\cdot \prod_{\beta\in \layers} \sum_{k\in\mathbb{N}} \pout(k) ((1-v)+vx_\beta)^{k},
\end{align}
which no longer depends on $\alpha$.
Therefore, the $G_\alpha$ are all equal, and since we begin the fixed point iteration with all $x_\alpha$ being equal as well,
the $x_\alpha$ will remain equal at each step of the iteration.
Thus also the components $q_\alpha$ of the fixed point will be equal.
We may therefore simplify the situation to a single generating function $G:\mathbb{R}\to\mathbb{R}$ that we iterate to arrive at a fixed point $q\in\mathbb{R}$.
We can see that $G$ is given by
\begin{align}
    G(x) &= \prod_{\beta\in \layers} \sum_{k\in\mathbb{N}} \pout(k) ((1-v)+vx)^k\\
         &= \gout((1-v)+vx)^{L},
\end{align}
where $\gout$ is once again the probability generating function of the out-degree distribution $\pout$.

\subsection{Cascade Condition from Probability}\label{apx-cond-prob-iid}
We substitute
\begin{equation}
    P(J,K)=\prod_{\alpha\in \layers} \pin(J_\alpha)\pout(K_\alpha)
\end{equation}
into
\begin{equation}
     M_{\alpha\beta} = \frac{1}{v z} \sum_{\overset{J,K}{J_\alpha=1}} P(J,K) \cdot K_\beta v_\beta,
\end{equation}
and find
\begin{align}
M_{\alpha\beta}
    &= \frac{1}{vz} \sum_{\overset{J,K}{J_\alpha=1}} \prod_{\gamma\in \layers} \pin(J_\gamma) \pout(K_\gamma) \cdot K_\beta v\\
    &= \frac{1}{z} \sum_{\overset{J,K}{J_\alpha=1}} \prod_{\gamma\in \layers} \pin(J_\gamma) \pout(K_\gamma) \cdot K_\beta.
\end{align}
This is the same as \eqref{eq-apx-cond-iid-1}, except that $\alpha$ and $\beta$ have switched places. The calculation thus proceeds as in section \ref{apx-cond-iid} and we obtain that $M_{\alpha\beta}=\pin(1)$ for all $\alpha,\beta\in\layers$.


\section{Equivalence of Cascade Conditions}\label{apx-proof-equivalence}
As seen in the main text, the two cascade conditions derived in sections \ref{sec-size-iid} and \ref{sec-prob-iid} agree in the special case of networks with i.i.d.~degrees.
A similar agreement of differently-derived cascade conditions has also previously been observed in another model of cascades \cite{Yagan2012}.
In this section, we prove that the two conditions are equivalent in general.

In order for both conditions to be defined, we need the assumptions of section \ref{sec-prob} to be satisfied.
So for this section, let $\hat L$ and $\hat P$ be as in section \ref{sec-prob} and assume that the network restricted to the layers $\nicelayers$ results in a branching process that is nonsingular and positively regular.

Let $\lambda_M$ be the leading eigenvalue of the matrix $M\in\mathbb{R}^{\hat L\times\hat L}$ given by
\begin{equation}
M_{\alpha\beta} = \frac{1}{v_\alpha z_\alpha} \sum_{\overset{J,K\in\mathbb{N}^{\hat L}}{J_\alpha=1}} \hat P(J,K) \cdot K_\beta v_\beta,
\end{equation}
and let $\lambda_A$ be the leading eigenvalue of the matrix $A\in\mathbb{R}^{L\times L}$ given by
\begin{equation}
    A_{\alpha\beta} = \frac{1}{z_\alpha} \sum_{\overset{J,K\in\mathbb{N}^L}{J_\beta=1}} P(J,K) K_\alpha
\end{equation}
Our goal is to prove that $\lambda_M > 1$ if and only if $\lambda_A > 1$.

We first address the fact that the matrices may have different dimensions, since $\hat L \le L$.
Consider $\beta\in \layers \setminus \nicelayers$.
From the definition of $\nicelayers$ it follows that
$\sum_{\overset{J,K}{J_\beta=1}} P(J,K) = 0$
and thus $A_{\alpha\beta}$ = 0 for all $\alpha\in \layers$.
In other words, $A$ has a zero column for every element of $\layers \setminus \nicelayers$.

We now drop these dimensions from $A$ and consider the submatrix $\hat A \in\mathbb{R}^{\hat L\times\hat L}$, defined by $\hat A_{\alpha\beta} \defeq A_{\alpha\beta}$.
The spectrum of $\hat A$ contains all nonzero eigenvalues of $A$.
We observe that
\begin{equation}
    v_\alpha z_\alpha \cdot M_{\alpha\beta} = v_\beta z_\beta \cdot \left( \hat A^\top  \right)_{\alpha\beta}.
\end{equation}
Let $D \in \mathbb{R}^{\hat L\times\hat L}$ be the diagonal matrix with $D_{\alpha\alpha} \defeq v_\alpha z_\alpha$.
Since these entries are always positive, $D$ is invertible.
Now we can restate the above equation as
\begin{equation}
    DM D^{-1} = \hat A^\top,
\end{equation}
meaning that $M$ and $\hat A^\top$ are similar and thus $M$ and $\hat A$ have the same spectrum.
It follows that all nonzero eigenvalues of $M$ and $A$ are the same and the cascade conditions are equivalent.


\section{Derivations: Constrained Multiplex Networks}
\label{apx-constrained-derivations}
The degree distribution of a constrained multiplex network is given by
\begin{equation}
\begin{split}
P(J,K) = \frac{1}{L} \sum_{\gamma\in\layers}
    \Bigg[
        \pout^{Lz_\gamma}(K_\gamma)
        &\left( \prod_{\beta\ne\gamma}\delta_0(K_\beta) \right)\\
        \cdot &\left( \prod_{\beta\in\layers} (1-C_{\beta\gamma})\delta_0(J_\beta) + C_{\beta\gamma}\pin(J_\beta) \right)
        \Bigg].
\end{split}
\end{equation}
For brevity, define
\begin{equation}
    f_{\gamma}(K) \defeq \pout^{Lz_\gamma}(K_\gamma) \prod_{\beta\ne\gamma}\delta_0(K_\beta)
\end{equation}
and
\begin{equation}
    g_{\gamma}(J) \defeq \prod_{\beta\in\layers} (1-C_{\beta\gamma})\delta_0(J_\beta) + C_{\beta\gamma}\pin(J_\beta).
\end{equation}

\subsection{Cascade Size}
\label{apx-size-constrained}
We fix $\alpha\in\layers$ and plug the degree distribution into the $\alpha$-component of the iterated map for the cascade size:
\begin{align}
q_\alpha(t+1)&=1- \frac{1-\rho_0}{z_\alpha} \sum_{J,K} P(J,K) K_\alpha \prod_{\overset{\beta\in \layers}{J_\beta>0}}\left( 1- q_\beta(t)^{J_\beta} \right)
\\&=
1- \frac{1-\rho_0}{Lz_\alpha} \sum_{J,K} \sum_{\gamma\in\layers} K_\alpha f_\gamma(K) \cdot g_\gamma(J)
\prod_{\overset{\beta\in \layers}{J_\beta>0}}\left( 1- q_\beta(t)^{J_\beta} \right)
\end{align}
To write this more compactly, we use
\begin{align}
g_\gamma(J)\prod_{\overset{\beta\in \layers}{J_\beta>0}}\left( 1- q_\beta(t)^{J_\beta} \right)
=&
\prod_{\beta\in\layers}
\begin{cases}
    \left(1-C_{\beta\gamma}\right) + C_{\beta\gamma} \pin(0) & \text{if } J_\beta=0 \\
    C_{\beta\gamma}\pin(J_\beta)\cdot\left(1-q_\beta(t)^{J_\beta}\right) & \text{if } J_\beta\ne 0
\end{cases}
\\ \eqdef &\prod_{\beta\in\layers} h_{\gamma\beta}(J_\beta).
\end{align}
Note that $\gamma\ne\alpha$ implies $K_\alpha f_\gamma(K) = 0$, and proceed:
\begin{align}
&1- \frac{1-\rho_0}{Lz_\alpha} \sum_{J,K} K_\alpha f_\alpha(K)
\cdot \prod_{\beta\in\layers}h_{\alpha\beta}(J_\beta)
\\={}&
1- \frac{1-\rho_0}{Lz_\alpha} \left(\sum_{K\in\mathbb{N}^L} K_\alpha
    \pout^{Lz_\alpha}(K_\alpha) \prod_{\beta\ne\alpha}\delta_0(K_\beta)\right)
\cdot \sum_J \prod_{\beta\in\layers}h_{\alpha\beta}(J_\beta)
\\={}&
1- \frac{1-\rho_0}{Lz_\alpha} \sum_{k\in\mathbb{N}} k \pout^{Lz_\alpha}(k)
\cdot \sum_J  \prod_{\beta\in\layers}h_{\alpha\beta}(J_\beta)
\\={}&
1- \frac{1-\rho_0}{Lz_\alpha} Lz_\alpha
\cdot \sum_J  \prod_{\beta\in\layers}h_{\alpha\beta}(J_\beta)
\end{align}
We apply Lemma \ref{lemma-sumprod} to obtain
\begin{align}
&1-(1-\rho_0) \sum_{J\in\mathbb{N}^L}  \prod_{\beta\in\layers} h_{\alpha\beta}(J_\beta)
\\={}&
1-(1-\rho_0) \prod_{\beta\in\layers} \sum_{j\in\mathbb{N}}  h_{\alpha\beta}(j)
\\={}&
1-(1-\rho_0) \prod_{\beta\in\layers} \left(
1-C_{\beta\alpha} + C_{\beta\alpha} \pin(0)
+ \sum_{j>0}  C_{\beta\alpha}\pin(j)\cdot\left(1-q_\beta(t)^{j}\right) \right)
\\={}&
1-(1-\rho_0) \prod_{\beta\in\layers} \left(
1-C_{\beta\alpha} + C_{\beta\alpha} \sum_{j\ge 0} \pin(j)
- C_{\beta\alpha} \sum_{j>0} \pin(j) q_\beta(t)^{j} \right).
\end{align}
Since $\pin$ is a probability distribution, we have $\sum_j \pin(j)=1$.
Furthermore, let $\gin(x)\defeq\sum_j\pin(j)x^j$ be
the probability generating function of $\pin$ and finally obtain
\begin{equation}
1-(1-\rho_0) \prod_{\beta\in\layers}
    1 + C_{\beta\alpha} \Big( \gin(0) - \gin(q_\beta(t)) \Big).
\end{equation}

\subsection{Cascade Probability}
Fix $\alpha$.
First, the probability of a link being vulnerable:
\begin{align}
v_\alpha&=\frac{1}{z_\alpha} \sum_{\overset{J,K}{J_\alpha=1}} P(J,K)
\\&=
\frac{1}{L z_\alpha}\sum_{\overset{J,K}{J_\alpha=1}}\sum_\gamma f_\gamma(K) g_\gamma(J)
\\&=
\frac{1}{L z_\alpha} \sum_\gamma \sum_{\overset{J}{J_\alpha=1}} g_\gamma(J) \sum_K f_\gamma(K).
\end{align}
Note that $\sum_K f_\gamma(K)=1$, as $f_\gamma(K)$ is the probability mass function of the out-degree distribution of a type-$\gamma$ node.
Also note that $\sum_{\overset{J}{J_\alpha=1}} g_\gamma(J) = \pin(1)C_{\alpha\gamma}$ is the probability that a type-$\gamma$ node is vulnerable on layer $\alpha$.
Obtain
\begin{align}
&\frac{1}{L z_\alpha} \sum_\gamma \pin(1)C_{\alpha\gamma}
\\={}&
\frac{\pin(1)}{L z_\alpha} \sum_\gamma C_{\alpha\gamma}.
\end{align}
Recall that $z_\alpha = \frac{\zin}{L} \sum_\beta C_{\alpha\beta}$, leading us to
\begin{align}
&\frac{\pin(1)}{L z_\alpha} \sum_\gamma C_{\alpha\gamma}
\\={}&
\frac{\pin(1)}{\zin}.
\end{align}
We now tackle the offspring generating function and find
\begin{align}
G_\alpha(\boldsymbol{x})&= \frac{1}{v_\alpha z_\alpha} \sum_{\overset{J,K}{J_\alpha=1}} P(J,K) \prod_{\beta} \big( (1-v_\beta) +v_\beta x_\beta \big)^{K_\beta}
\\&=
\frac{1}{L v_\alpha z_\alpha} \sum_\gamma
\left[ \sum_K f_\gamma(K) \prod_{\beta} \big( (1-v_\beta) +v_\beta x_\beta \big)^{K_\beta} \right]
\sum_{\overset{J}{J_\alpha=1}} g_\gamma(J)
\\&=
\frac{1}{L v_\alpha z_\alpha} \sum_\gamma
\left[ \sum_{k\in\mathbb{N}} \pout^{Lz_\gamma}(k) \big( (1-v_\gamma) +v_\gamma x_\gamma9 \big)^k \right]
\pin(1)C_{\alpha\gamma}
\\&=
\left( \sum_\beta C_{\alpha\beta} \right)^{-1} \sum_\gamma
\gout^{Lz_\gamma}\big( (1-v_\gamma) +v_\gamma x_\gamma \big)
\cdot C_{\alpha\gamma},
\end{align}
where $\gout^{L z_\gamma}$ is the probability generating function of $\pout^{L z_\gamma}$.


\section{Explosive Onset: Calculation of \texorpdfstring{$c_{xx}$}{c\_xx}}
\label{apx-explosive-cxx}
This calculation follows the method outlined in section 5.4 of \cite{Kuznetsov2023}.\\
We have eigenvector $q=\pmat{1&1&1}^\top$ and adjoint eigenvector $\hat q = \frac{1}{2+p}\pmat{1&1&p}^\top$, normalized so that $\langle\hat q,q\rangle = 1$.
We fix $z$ to the critical value $z^\star(p)$ and drop it from the notation of $F$, i.e.~we write $F(x)\defeq F(x,z^\star(p))$. We then expand $F$ in $x$ as
\begin{equation}
    F(x) = Ax + \frac{1}{2} B(x,x) + O(\lVert x \rVert^3),
\end{equation}
where $A$ is the Jacobian of $F$ at $\mathbf{0}$ and $B:\mathbb{R}^3\times\mathbb{R}^3\to\mathbb{R}^3$ is the multilinear function given by
\begin{equation}
    B_\alpha(x,y) = \sum_{\beta,\gamma} \frac{\mathrm d^2F_\alpha(\xi)}{\mathrm d\xi_\beta \mathrm d\xi_\gamma} \bigg\rvert_{\xi=0} x_\beta y_\gamma.
\end{equation}
We parameterize the center manifold as
\begin{equation}
    H(\xi) = \xi q + \frac{1}{2} \xi^2 h_2 + O(\xi^3),
\end{equation}
where $h_2\in\mathbb{R}^3$ is some vector whose value we won't need explicitly.
Restricting $F$ to the center manifold yields a map
\begin{equation}
    G(\xi) = \xi + c_{xx}\xi^2 + O(\xi^3),
\end{equation}
which satisfies
\begin{equation}
    F \circ H = H \circ G.
\end{equation}
We expand both sides of the above equation to
\begin{align}
F(H(\xi))
    &= A(\xi q + \frac{1}{2}\xi^2h_2+\cdots) 
    + \frac{1}{2}B(\xi q + \cdots, \xi q + \cdots)
    + \cdots\\
H(G(\xi))
    &= q(\xi+c_{xx}\xi^2+\cdots)
    + \frac{1}{2}h_2(\xi+\cdots)^2
    +\cdots.
\end{align}
Comparing the $\xi^2$ terms yields
\begin{equation}
    \frac{1}{2}Ah_2 + \frac{1}{2} B(q,q) = c_{xx} q + \frac{1}{2} h_2
\end{equation}
and thus
\begin{equation}
    (\mathrm{Id} - A) h_2 = B(q,q) - 2c_{xx}q.
\end{equation}
We have
\begin{align}
     & \langle \hat q, (\mathrm{Id} - A) h_2 \rangle\\
    ={}& \langle \hat q, h_2 \rangle - \langle A^\top \hat q, h_2\rangle\\
    ={}& 0,
\end{align}
which implies
\begin{align}
0   ={}& \langle \hat q, (\mathrm{Id} - A) h_2 \rangle\\
    ={}& \langle \hat q, B(q,q)-2c_{xx}q\rangle\\
    ={}& \langle \hat q, B(q,q)\rangle - 2c_{xx}
\end{align}
and thus
\begin{equation}
    c_{xx} = \frac{1}{2} \langle \hat q, B(q,q) \rangle.
\end{equation}
At this point we evaluate
\begin{align}
    B_1(q,q) &= \gin^{\prime\prime}(0) - 2\gin^\prime(0)^2\\
    B_2(q,q) &= \gin^{\prime\prime}(0) - 2\gin^\prime(0)^2\\
    B_3(q,q) &= p\gin^{\prime\prime}(0) - 2p^2\gin^\prime(0)^2,
\end{align}
where we recall $\gin(s)=e^{z^\star(p)(s-1)}$.
We note that $\gin^\prime(0) = (2+p)^{-1}$ and $\gin^{\prime\prime}(0)=z^\star(p)(2+p)^{-1}$ and compute
\begin{align}
c_{xx}
    &= \frac{1}{2} \cdot \frac{1}{2+p}
        \cdot \Big( 2\left( \gin^{\prime\prime}(0) -2 \gin^\prime(0)^2 \right)
                   - p\left(p\gin^{\prime\prime}(0) -2p\gin^\prime(0)^2\right)
                \Big)\\
    &= \frac{1}{2} \cdot \frac{1}{2+p}
        \cdot \Big( \left(2+p^2\right) \gin^{\prime\prime}(0)
                   -\left(4+2p^3\right)   \gin^\prime(0)^2
                \Big)\\
    &= \frac{1}{2} \cdot \frac{1}{2+p}
        \cdot \left( \frac{2+p^2}{2+p} z^\star(p)
                   -2\cdot\frac{2+p^3}{(2+p)^2}
                \right)\\
    &=  \frac{1}{(2+p)^2}\cdot\left(
            \frac{1}{2} \left(2+p^2\right) \cdot z^\star(p) - \frac{2+p^3}{2+p}
            \right).
\end{align}


\section{Comparison to Flattened Model}
\label{apx-projected-theory}
To show that the phenomena presented in section \ref{sec-constraint-examples} truly require a multiplex formulation of the model, we analyze a flattened version of the Constrained Multiplex Network Model.
Specifically, we derive the cascade size for the classic Watts model on a monoplex configuration model whose degree distribution is the total (i.e.~flattened) degree distribution of a given CMNM.
The thresholds are chosen to match the typical number of neighbors needed to activate a node.

Assume we are given a Constrained Network Model with a parameter $z\in(0,\infty)$, $\pin=\Pois(z)$, $(\pout^{\hat z})_{\hat z>0} = (\Pois(\hat z))_{\hat z>0}$, and $C \in \{0,1\}^L$ (not $[0,1]^L$).
We define the \textit{in-participation} of a node type $\alpha\in\layers$ as $\inpar{\alpha} \defeq \sum_\beta C_{\beta\alpha}$, and the \textit{out-participation} as $\oupar{\alpha} \defeq \sum_\beta C_{\alpha\beta}$.
Then we can write the total degree distribution as
\begin{equation}
    P(j,k) = \frac{1}{L} \sum_\alpha \pois{z \inpar{\alpha}}{j} \cdot \pois{z \oupar{\alpha}}{k},
\end{equation}
where $\pois{\lambda}{k}$ denotes the probability mass function of $\Pois(\lambda)$.

Since the different node types in the CMNM can be more or less robust, we let the thresholds depend on $\alpha$, thus making them correlated with the degrees.
Specifically, we choose
\begin{align}
    \phi_\alpha &\defeq \frac{1}{\text{total \# in-links}}\left(\frac{\sum_\beta \text{\# in-links on layer } \beta}{\text{\# layers with in-links}} \right)\\
    & = \frac{1}{\inpar{\alpha}}.
\end{align}

To derive the cascade size $\rho(t)$, we proceed similarly as in section \ref{sec-size}.
Let $q(t)$ denote the probability that a randomly chosen link is active at time $t$.
Fix a focal node $v$. With probability $\rho_0$, $v$ is a seed node.
Otherwise, $v$ could have been activated with probability
\begin{equation}
    \frac{1}{L} \sum_\alpha \sum_{j \ge 1} \pois{z\inpar{\alpha}}{j} \cdot B(j,q(t);j\phi_\alpha),
\end{equation}
where
\begin{equation}
    B(n,p;x) \defeq 1-\sum_{k=0}^{\lceil x-1 \rceil}\bin(n,p;k),
\end{equation}
and $\bin(n,p;k)$ is the probability mass function of the binomial distribution.
Intuitively, $B(n,p;x)$ is the probability of getting $x$ or more successes in $n$ coin tosses with success probability $p$.

Altogether, we obtain
\begin{equation}
    \rho(t) = \rho_0 + (1-\rho_0)\cdot \frac{1}{L} \sum_\alpha \sum_{j \ge 1} \pois{z\inpar{\alpha}}{j} \cdot B(j,q(t);j\phi_\alpha)
\end{equation}
and similarly
\begin{equation}
    q(t) = \rho_0 + (1-\rho_0)\cdot \sum_\alpha \frac{\oupar{\alpha}}{\sum_\beta\oupar{\beta}} \sum_{j \ge 1} \pois{z\inpar{\alpha}}{j} \cdot B(j,q(t);j\phi_\alpha).
\end{equation}
The factor $\frac{\oupar{\alpha}}{\sum_\beta\oupar{\beta}}$ replaces $\frac{1}{L}$ because a randomly chosen link is more likely to originate from a high-out-degree node.
We will refer to this as the \textit{flattened model}.

Regarding the restriction that all $C_{\alpha\beta}\in\{0,1\}$, note that this is necessary for a closed form of the total degree distribution, and for the motivation behind $\phi_\alpha$.
But as we lack better options, we will still use the flattened model if there are $C_{\alpha\beta}\in(0,1)$.

\begin{figure}
    \centering
    \includegraphics[width=0.7\linewidth]{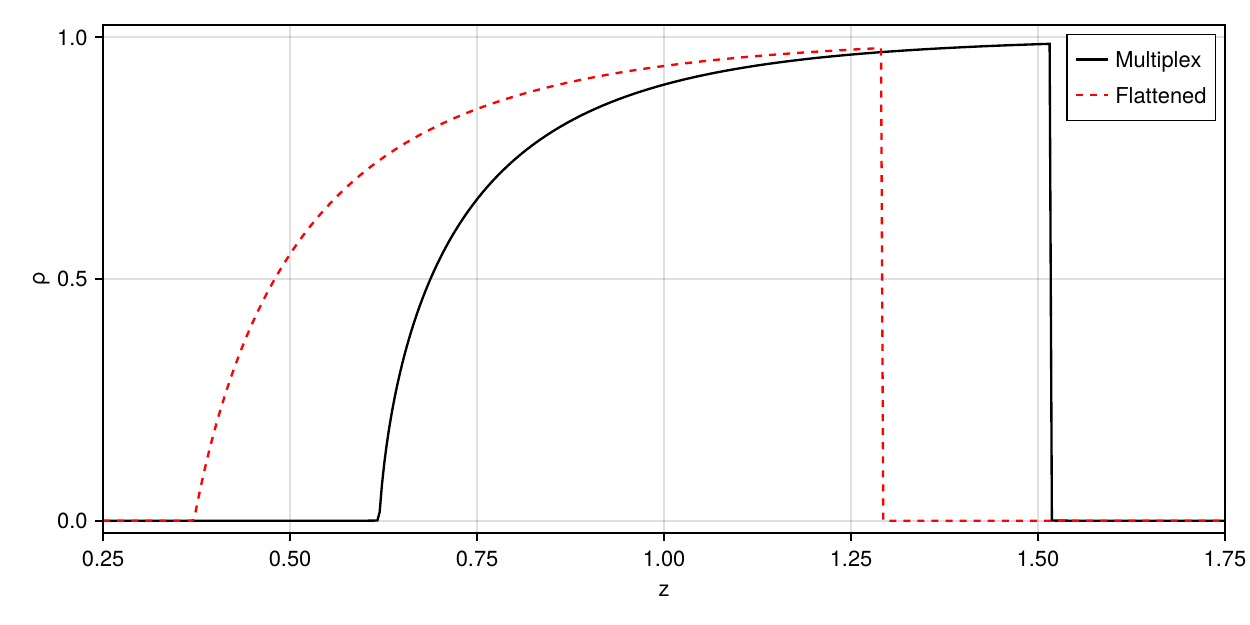}
    \caption{Analytic cascade sizes in the multiplex model vs.~the flattened model.}
    \label{fig:proj-ones33}
\end{figure}

\begin{figure}
    \centering
    \subfloat{
    \begin{overpic}[width=0.7\linewidth]{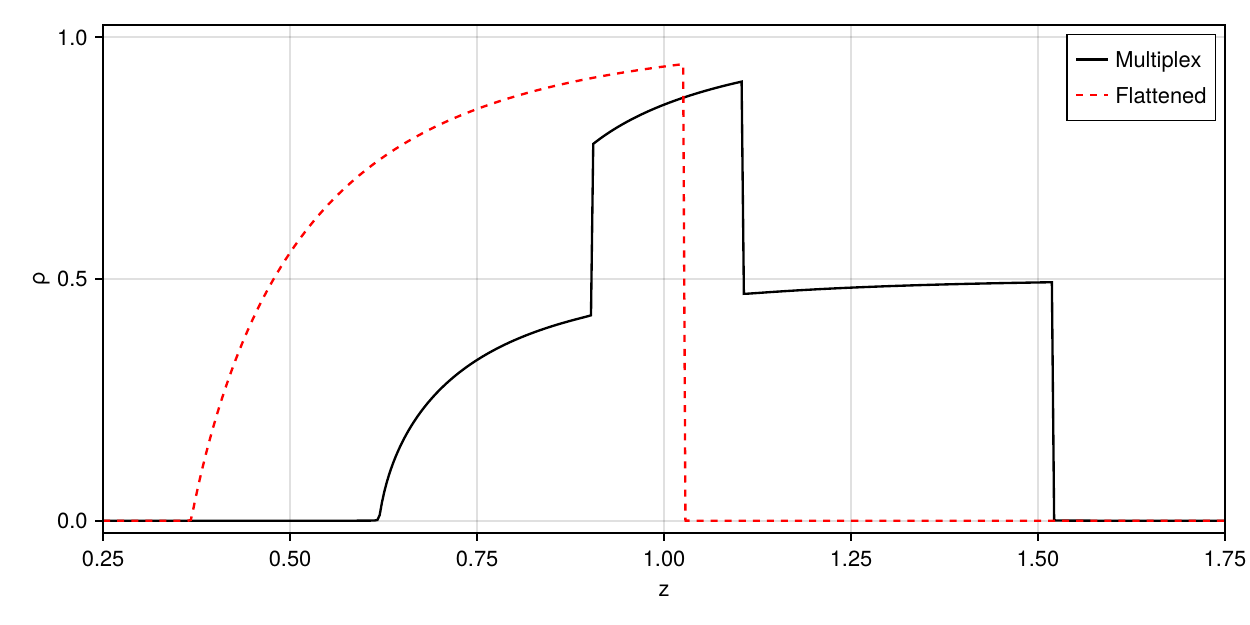}
        \put(9,43){(a)}
    \end{overpic}}\\
    \subfloat{
    \begin{overpic}[width=0.7\linewidth]{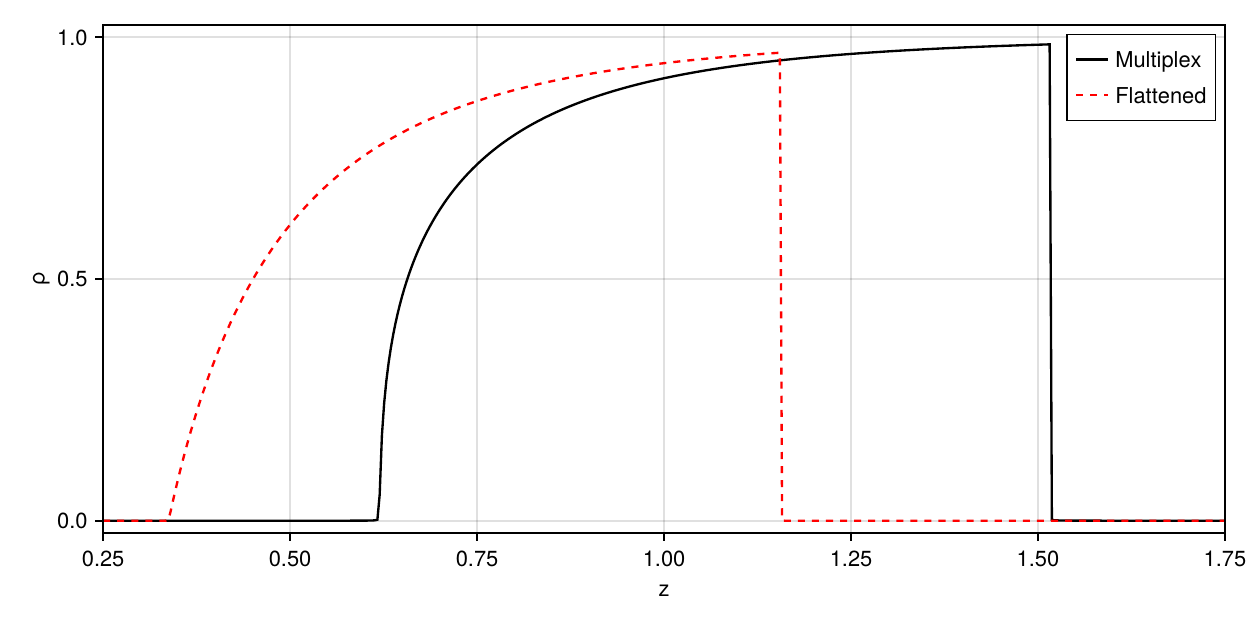}
        \put(9,43){(b)}
    \end{overpic}}\\
    \subfloat{
    \begin{overpic}[width=0.7\linewidth]{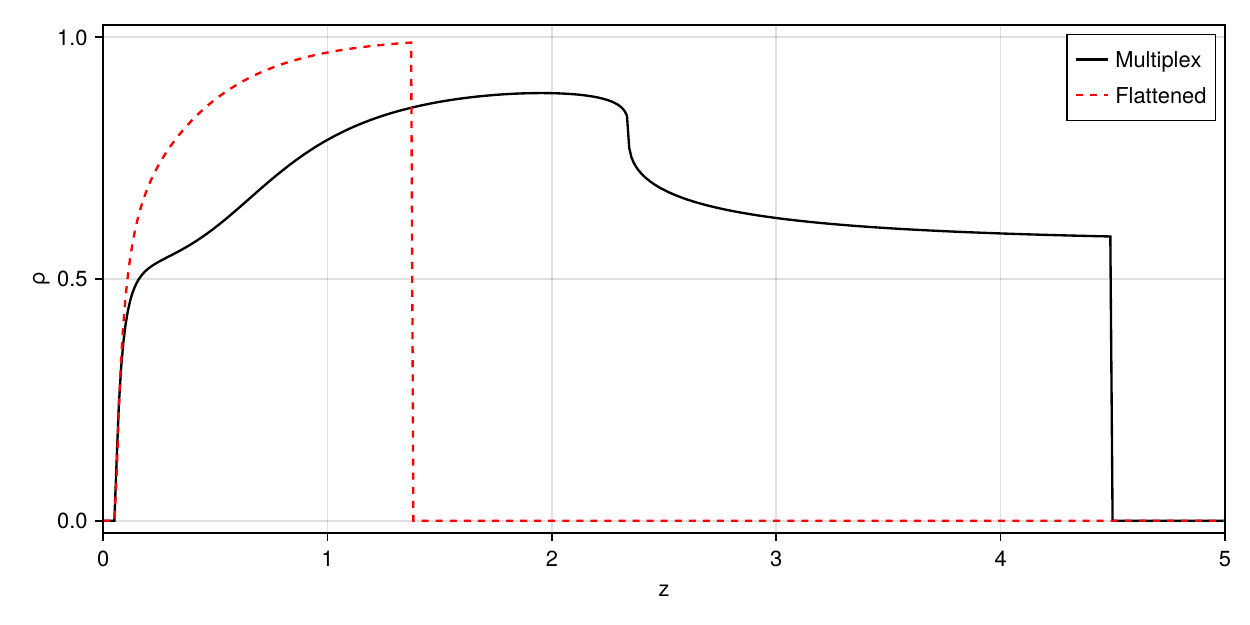}
        \put(9,43){(c)}
    \end{overpic}}
    \caption{Analytic cascade sizes in the multiplex model vs.~the flattened model.
        (a) $C$ as in Fig.~\ref{fig:nested-a-micro}.
        (b) $C$ as in Fig.~\ref{fig:nested-b}.
        (c) $C$ as in Fig.~\ref{fig:twoish-micro}.}
    \label{fig:proj-nontrivial}
\end{figure}

We now compare the cascade size predicted by the flattened model to the formula derived in section \ref{sec-size}, which - as seen in section \ref{sec-constraint-examples} - matches simulations of the multiplex model.
First, we choose the simplest nontrivial constraint matrix
\begin{equation}
    C = \pmat{1&1&1\\1&1&1\\1&1&1}.
\end{equation}
The result is shown in Fig.~\ref{fig:proj-ones33}.
We see that both models predict qualitatively the same behavior, but the flattened model does not predict the correct placement of the phase transitions.
If we instead choose the constraint matrices presented in section \ref{sec-constraint-examples}, the agreement becomes much worse, as shown in Fig.~\ref{fig:proj-nontrivial}.
We conclude that the flattened model presented here completely fails to capture the behavior of the multiplex model.

\end{document}